\documentclass[pdflatex,sn-chicago]{sn-jnl}

\usepackage{array, graphicx}%
\usepackage{multirow}%
\usepackage{amsmath,amssymb,amsfonts}%
\usepackage{amsthm}%
\usepackage{mathrsfs}%
\usepackage[title]{appendix}%
\usepackage{xcolor}%
\usepackage{textcomp}%
\usepackage{manyfoot}%
\usepackage{booktabs}%
\newcommand{\ra}[1]{\renewcommand{\arraystretch}{#1}}
\usepackage{listings}%
\usepackage{graphicx}
\usepackage{caption}
\usepackage{subcaption}
\usepackage{lmodern}
\usepackage{tabularx,ragged2e}
\usepackage{appendix}
\usepackage{lscape} 
\usepackage{bm}
\usepackage{subcaption}
\usepackage{multirow}
\usepackage{arydshln}
\usepackage{blindtext}
\usepackage{xpatch}

\begin{document}

\title[Article Title]{Twenty-Four Years of Empirical Research on Trust in AI: A Bibliometric Review of Trends, Overlooked Issues, and Future Directions}

\author*[1,2]{\fnm{Michaela} \sur{Benk}}\email{mbenk@ethz.ch}

\author[2]{\fnm{Sophie} \sur{Kerstan}}

\author[2]{\fnm{Florian} \sur{von Wangenheim}}

\author[3]{\fnm{Andrea} \sur{Ferrario}}

\affil*[1]{\orgdiv{Mobiliar Lab for Analytics}, \orgname{ETH Zurich}, \orgaddress{\city{Zurich}, \country{Switzerland}}}

\affil[2]{\orgdiv{Department of Management, Technology, and Economics}, \orgname{ETH Zurich}, \orgaddress{\city{Zurich}, \country{Switzerland}}}

\affil[3]{\orgdiv{Institute of Biomedical Ethics and History of Medicine}, \orgname{University of Zurich}, \orgaddress{\city{Zurich}, \country{Switzerland}}}

\abstract{Trust is widely regarded as a critical component to building artificial intelligence (AI) systems that people will use and safely rely upon. As research in this area continues to evolve, it becomes imperative that the research community synchronizes its empirical efforts and aligns on the path toward effective knowledge creation. To lay the groundwork toward achieving this objective, we performed a comprehensive bibliometric analysis, supplemented with a qualitative content analysis of over two decades of empirical research measuring trust in AI, comprising 1'156 core articles and 36'306 cited articles across multiple disciplines. Our analysis reveals several ``elephants in the room'' pertaining to missing perspectives in global discussions on trust in AI, a lack of contextualized theoretical models and a reliance on exploratory methodologies. We highlight strategies for the empirical research community that are aimed at fostering an in-depth understanding of trust in AI.}


\maketitle

\newpage
\section{Introduction}\label{introduction}
Trust in artificial intelligence (AI)-based systems is a focal concern among researchers and practitioners alike, and increasingly so. At the time of writing, a search for ``trust in AI'' on Google Scholar returns 157 results for all years until 2017, 1’140 results for the years 2018-2020, and 7'310 results for the years 2021-2023. The surge of interest in how humans trust AI systems is fueled by increasingly capable AI-based technologies that are transforming the way humans work and collaborate in a multitude of society-impacting applications \citep{Jacobs2021DesigningAF, Chiou2021TrustingAD, Lucaj2023AIRI, Shin2021TheEO, Wamba2020AreWP, Feher2024ModelingAT}. 
 
This emphasis on trust builds upon a substantial legacy of prior research focused on trust in technology more broadly \citep{lee_trust_2004, mcknight_chervany_1996}, interpersonal trust \citep{mayer-et_al_1995, Castelfranchi_Falcone_2010}, or trust in institutions \citep{Knowles2021TheSO, Kroeger2016FaceworkCT}. Whereas in organizations and interpersonal contexts, trust is considered the cornerstone of effective communication, collaboration in teams, and healthy and positive work cultures \citep{mayer-et_al_1995}, in the context of human interactions with technology more broadly, trust is widely regarded as instrumental to enhance effectiveness and efficiency. This is underpinned by a range of desired outcomes of trust, such as successful adoption, acceptance, and appropriate use \citep{Jacobs2021DesigningAF, Chiou2021TrustingAD, Shin2021TheEO, Buccinca_2021}. As AI increasingly blurs the distinction between tool and teammate \citep{Rix2022FromTT}, a fast-growing research community continues to conduct empirical investigations on trust in AI systems, while engaging in an ongoing discourse on the unique issues and pitfalls surrounding empirical practices and measurement of trust in AI research \citep{Buccinca2020ProxyTA, Jacovi2020FormalizingTI, Schraagen2020TrustingTX, Chi2021DevelopingAF, schemmer2022should}. To streamline these discussions, leading academic journals and conferences have introduced initiatives dedicated to the topic. They include ``CHI TRAIT'' \citep{TRAIT} at the ACM Conference on Human Factors in Computing Systems (CHI) or ``HRI TRAITS,''\footnote{https://sites.google.com/view/traits-hri-2022} at the ACM/IEEE International Conference on Human-Robot Interaction (HRI). Additionally, Computers in Human Behavior recently organized a special issue on the topic.\footnote{https://www.sciencedirect.com/special-issue/102LJP9BFH9} Nevertheless, research in this area remains fragmented, in part due to its conceptual heterogeneity \citep{Laux2023TrustworthyAI}. This fragmentation complicates the comparison of findings across various research domains, presenting a significant challenge to practitioners and policymakers.

The rapid growth of interdisciplinary research on trust in AI offers an opportunity to reflect on this field and place it within a broader historical context. Analyzing publication patterns and trends can help map the development of this research over time and address current measurement issues within this broader perspective. Moreover, by characterizing its publication patterns and knowledge structure across research domains and over time, researchers could better understand which research trends impede or foster progress in our understanding of trust in human-AI interactions, align research efforts, ultimately leading to the design of AI systems that humans are confident to use and rely upon effectively.
 \citep{kaur2022trustworthy}. To facilitate this reflection, this work aims to provide a systematic overview of the empirical research of trust in AI through a comprehensive mapping of over two decades of scientific output. To this end, we employed a mixed-method approach, consisting of bibliometric and content analyses of empirical research measuring trust in AI, thereby shedding light on multiple research trends. Specifically, our contributions are as follows:

\begin{itemize}

\item Based on a comprehensive bibliometric analysis of over two decades of empirical research on trust in AI, encompassing 1’156 core articles and 36’306 cited articles across various research domains, we uncover publication patterns, describe the underlying knowledge structure of the field, and examine the most influential articles within each domain. To enrich the quantitative insights, we further conducted a qualitative content analysis of the top 10 \% most cited articles per domain. 
\item We identify and challenge under-addressed issues within the field, prompting critical questions about these ``elephants in the room.'' These issues pertain to (a) the research production that is shaped by Western and Global Northern perspectives and driven by technical domains; (b) the rapid diversification of application areas for AI, which hampers in-depth exploration of the concept of trust in AI; (c) the predominantly exploratory research approach that still characterizes its scientific output.
\item We offer practical recommendations that aim to address these limitations by promoting the inclusivity of research on trust in AI, advocating for the development and application of contextualized frameworks, and encouraging systematic investigations to deepen our understanding and measurement of trust in AI.

\end{itemize} 

Our results and recommendations have the potential to significantly influence the future direction of empirical research on trust in AI by encouraging scientific practices to achieve the theoretical clarity and methodological robustness this field finally deserves. This, in turn, will facilitate the development of theory-grounded and evidence-based policies that effectively foster the social good we commonly refer to as trust in AI.

This paper is structured as follows: In Section \ref{related-work}, we provide working definitions of the main concepts of relevance for empirical research on trust in AI. In particular, in Section \ref{reviews}, we provide an overview of the current state of knowledge on empirical research on trust in AI and introduce the research questions that guide this work. In Section \ref{methodology}, we provide an overview over the methodology used in this work, including the search strategy, article selection criteria, and methods used to analyze the data. In Section \ref{results}, we present the findings from our analyses, which we discuss in Section \ref{discussion}. Here, we highlight the ``elephants in the room'' and propose strategies for future investigations on trust in AI in Section \ref{strategies}.
Lastly, we summarize our work's limitations and conclusions in Sections \ref{limitations} and \ref{conclusion}, respectively. 

\section{Trust in AI: State of knowledge}
\label{related-work}

\subsection{Trust theory: A basis to agree on}
\label{trust_theory}

Trust emerges in situations where humans face uncertainty, risk, and interdependence \citep{McKnight2000WhatIT}. At its core, trust arises as a relational process involving a trustor and a trustee, situated within a specific context and driven by a specific goal \citep{Castelfranchi_Falcone_2010}. It is now a relevant research topic in human-machine interactions \citep{lee_trust_2004, mcknight_trust_2011}, prominently in the case of AI systems \citep{Jacovi2020FormalizingTI, Chiou2021TrustingAD, toreini_relationship_2019}. 
Despite different definitions appearing in the literature,\footnote{See e.g., Castaldo (2010) for a content analysis of 72 definitions of interpersonal trust, using 273 related terms \citep{Castaldo2010TheMO}. For the case of human-AI interactions, we refer to, for instance, \citep{Jacovi2020FormalizingTI, ferrario_ai_2019, ferrario2022explainability, loi2023much}} research agrees that trusting necessarily involves the \emph{trustor}, i.e., the trusting agent, the \emph{trustee}, i.e., the trusted agent and the goal and context of the trusting relation.\footnote{The context and the goal of trusting remind us that we do not usually trust others in every possible way. I may trust my experienced dentist to perform a root canal flawlessly, but not to perform a heart surgery. Likewise, I may trust my AI-based health coach to help me lowering cholesterol levels, but not suggesting dating advice.}  
In particular, there is a growing interest in defining what makes the trustee in an interaction, that is, the AI system, worthy of trust   \citep{jobin2019global,li2021trustworthy,petersen2022responsible,kaur2022trustworthy}. Authors 
argue that the trustworthiness of an AI system comprises different properties, such as reliability, robustness and transparency \citep{Jacovi2020FormalizingTI,ferrario2022explainability,kaur2022trustworthy}. Here, research suggests that trust should be grounded in the trustworthiness of the AI. By ensuring user trust is calibrated with the AI's capabilities, the risk that the user overtrusts and mistrusts the AI is mitigated\citep{lee_trust_2004,Jacovi2020FormalizingTI,ferrario2022explainability}. Finally, research often investigates trust in relation to other constructs, including reliance \citep{lee_trust_2004}. Reliance is the behaviour that the trustor performs after delegating a task to the trustee in order to achieve a goal. For instance, a physician relies on a medical AI by accepting the predictions computed by the system as the result of delegating the task of computing accurate predictions to the AI. As such, reliance can -- but does not have to -- follow trusting \citep{lee_trust_2004}. This concept is  applicable also in situations of ``reliance without trust'', effectively amounting to reliance accompanied by a minimal, or zero, degree of users' trust \citep{loi2023much}. Such instances underscore the distinction between trust and reliance, illustrating that while they are interconnected, they are fundamentally separate constructs.


\subsection{Reviews of empirical research on trust in AI}
\label{reviews}

A few works have contributed insights for empirical research on trust in AI by reviewing relevant literature. For instance, \citet{Glikson2020HumanTI} performed a literature review of ``human trust in AI'' (no focus on empirical methods) with a search on Google Scholar and, subsequently on Business Source Premier, Engineering Village, and PsycINFO, considering the years between 1999 and 2019. Targeting research in organizational science, the review analyzes the determinants of human trust in AI and describes how AI and its integration in organizations differs from other technologies \citep{Glikson2020HumanTI}. Reviewing 150 articles, the authors argue that the (a) representation (robot, virtual, and embedded) and (b) capabilities of an AI system play an important role in how human-AI trust develops \citep{Glikson2020HumanTI}. They further highlight the need to investigate long-term human-AI interactions, calling for more interdisciplinary collaboration to improve research methodologies \citep{Glikson2020HumanTI}. 

Furthermore, \citet{Vereschak2021HowTE} conducted a systematic review to examine how trust is measured in studies on AI-assisted decision-making, namely, in situations where ``humans make decisions based on their own expertise and on recommendations provided by an AI-based algorithm'' \cite[p.~327:1]{Vereschak2021HowTE}. They retrieved 83 articles from the period between 2005 and 2020 using the ACM Digital Library with a search based on keywords (e.g., ``artificial intelligence'' and ``trust'') that did not address specifically the empirical production on trust in AI \citep{Vereschak2021HowTE}. Their review shows that trust research employs different definitions that are adopted or adapted from the social sciences \citep{Vereschak2021HowTE}. However, these definitions are not appropriately integrated in empirical works, affecting the interpretation and comparison of results across studies \citep{Vereschak2021HowTE}, as also remarked by \citet{benk2022value}. The authors propose a number of guidelines to improve the empirical study of trust in AI-assisted decision-making. Examples are clarifying the chosen definition of trust in AI, differentiating it from other constructs (e.g., reliance), favouring the analysis of trust in AI in long-term, real-world human-AI interactions and the use of well-established instruments to measure it \citep{Vereschak2021HowTE}. 

Beyond these traditional reviews, \citet{knickrehm2023can} recently used topic modeling on the abstracts of publications on trust in AI retrieved from Scopus, Web of Science, and the AIS Library. They conducted a broad search on trust in AI, not focusing specifically on empirical research, by searching for the keywords ``trust'' and ``artificial intelligence'' in the abstracts of publications. Performing topic modeling with Bidirectional Encoder Representations from Transformers (BERT) \citep{devlin2018bert}, they uncovered 56 topics from the abstracts of 3'356 articles published between 1986 and 2022. These topics are then manually collected in 11 clusters, such as ``Governance'', ``Social Justice'' and ``Robot-Human-Interaction'' \citep{knickrehm2023can}. Further, the authors manually identified human-, organization- and AI-related factors in the abstracts, such as confidence, social responsibility and transparency, that influence trust in AI \citep{knickrehm2023can}. 

As highlighted in previous paragraphs, existing reviews of trust in AI have focused on (a) domain-specific aspects of trust (e.g., AI representation within organizational science); (b) empirical methods (e.g., definitions and key elements of experimental protocols); and (c) important factors of trust in AI more broadly. 
However, trust research continues to be hampered by fragmentation and conceptual heterogeneity \citep{Laux2023TrustworthyAI}, complicating the comparison of findings across different domains and posing a significant challenge to attempts to review the structure of the field.
Consequently, the current state of knowledge regarding trust in AI is insufficient to provide clear guidelines for research advancement. Our work addresses these limitations, as elaborated in the following sections.

\subsection{Goal of this work and research questions}
\label{RQS}
Previous related works have explored trust in AI, offering insights into its integration in organizations \citep{Glikson2020HumanTI}, sharing recommendations for empirical studies on AI-assisted decision-making \citep{Vereschak2021HowTE}, and providing an exploratory
mixed-method analysis of trust in AI using abstracts of relevant work \citep{knickrehm2023can}. Our work differs from theirs in several ways. First, we review empirical trust in AI literature across various research domains and contexts and do not limit ourselves to a particular domain or area of application. In particular, we do not restrict our analysis to the literature on specific AI systems, such as conversational AI or autonomous cars.
Second, we employ a mixed-method approach, integrating a bibliometric with a qualitative content analysis. Doing so, we can explore the landscape of empirical research on trust in AI as a \textit{research field} and characterize the influential works that shape its production over time mapping out the quantitative trends and patterns, as well as to delve deeper into the articles qualitatively.
Our goal is to thereby uncover the field's patterns, trends and gaps and, ultimately, provide a set of strategies for research that are applicable across the various domains and contexts in which trust in AI is empirically explored. Specifically, we address the following research questions:

\begin{enumerate}
\item[\textbf{RQ1}] Which publication patterns characterize the literature on empirical research on trust in AI? 
\item[\textbf{RQ2}] What are the conceptual themes and intellectual citation trends that shape the empirical research community's understanding and investigation of trust in AI?
\item[\textbf{RQ3}] How do the overarching patterns and trends manifest within empirical studies and shape investigations on trust in AI?
\end{enumerate}


\section{Methodology}
\label{methodology}

\subsection{Search strategy and query}\label{search}
We conducted a systematic search of peer-reviewed literature available from 2000 to 2023, using the electronic databases Scopus, Web of Science, ACM Library, PubMed, and IEEE Explorer.
Databases were chosen to broadly cover different interdisciplinary lenses on the measurement of trust in AI. The search used a combination of keywords and controlled vocabulary for the concepts of AI, trust and measurement. The search was independently piloted using Scopus and ACM Library by two investigators and subsequently adapted to each database. The identification and extraction of data for the bibliometric analysis conforms to the Preferred Reporting Items for Systematic Reviews and Meta-Analyses (PRISMA) protocol \citep{Page2020TheP2}. 
Following prior literature reviews on trust \citep{Vereschak2021HowTE, Glikson2020HumanTI,knickrehm2023can}--see Section \ref{reviews}, we considered all works that referred to their measurement construct as ``trust,'' thereby including those studies focusing on trusting attitudes, beliefs, intentions, and trusting behaviors, but excluding those that focus on related, but considered distinct constructs, such as ``reliance'' or ``disposition to trust.'' 

Furthermore, as our work attempts to capture the interdisciplinary nature of empirical research on trust in AI, we wanted to include a variety of terms describing AI systems in our search, but exclude technologies that we deemed out-of-scope of this work, such as automated machinery.  
Examples of terms that are used in the literature to describe AI systems are ``artificial intelligence,'', ``algorithm'', ``machine learning-based system'', or ``robot'' \citep{Langer2022}. Further,  different disciplines may prefer one term over another.   
Therefore, to mitigate disciplinary biases, a total of eight researchers from diverse disciplines (including Psychology, Information Systems, Marketing, Computer Science, and Human Computer Interactions) reviewed and refined the terminology used in the search. Our strategy was informed by the framework provided by \citet{Glikson2020HumanTI}, which includes three types of AI representations: robots, virtual, and embedded. Following the framework, we differentiate between automation and AI on the basis that traditional automation does not encompass learning processes, a key characteristic of AI. 
\\
Our final search query used a combination of terms related to \textit{AI}, \textit{trust}, and \textit{measurement}, resulting in the following query: 

\begin{lstlisting}[basicstyle=\small\sffamily, breaklines=true]
(``artificial intelligence'' OR ``intelligent agent'' OR  ``neural network'' OR ``deep learning'' OR ``machine learning'' OR ``learning algorithm'' OR robot* OR ``autonomous car'' OR ``autonomous vehicle'' OR  ``natural language processing'' OR ``recommender system'' OR ``ai'' OR ``xai'' OR ``expert system'') AND (trust*) AND (measure* OR experiment* OR empiric* OR assess* OR "questionnaire" OR (``survey'' AND NOT ``taxonomy'') OR ``user study'' OR (``human evaluation'' AND NOT (``trust network'' OR ``social trust'' OR ``taxonomy''))). 
\end{lstlisting}

The search included English publications from the years 2000 until 2023. The search was carried out in the spring of 2021. Two additional searches were carried out in the spring of 2022 for the year 2021, and in the spring of 2024 for the years 2022 and 2023.

\subsection{Article selection}\label{selection}
\begin{figure}[h!]
    \centering
    \includegraphics[width=10cm]{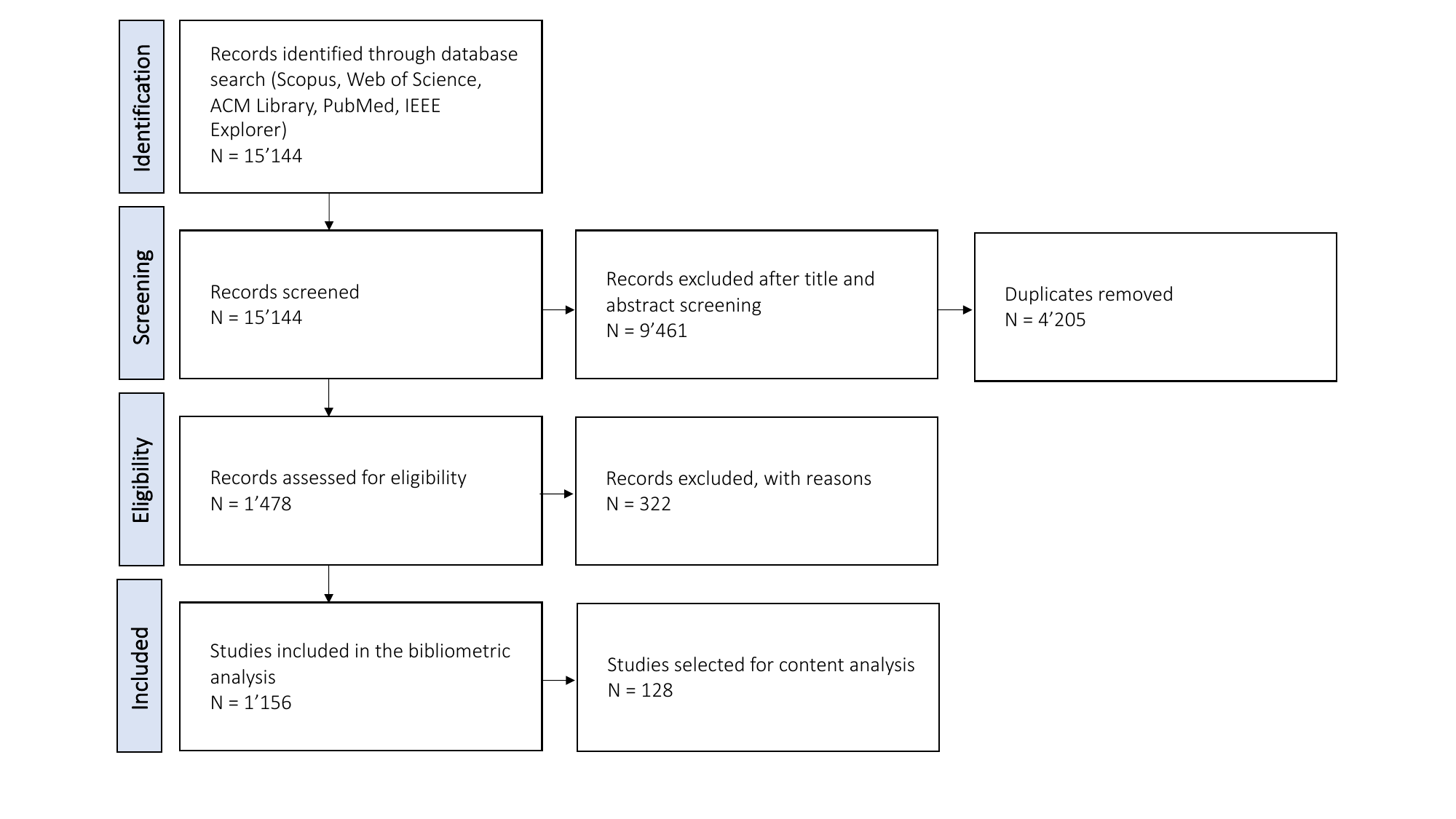}
    \caption{PRISMA flow diagram}
    \label{fig:prisma}
\end{figure}

Study selection occurred in three phases. First, two independent investigators conducted the initial screening of a total of 15'144 identified articles, based on titles and abstracts. Next, selected articles were exported into Mendeley, a reference management software, and duplicates removed. Finally, 1'478 full-text articles were screened, using a set of inclusion and exclusion criteria by two independent investigators. All studies were labeled by two independent investigators as meeting the criteria, not meeting the criteria, or possibly meeting the criteria. Disagreements were resolved through discussion until consensus was reached. At the conclusion of this phase, 1’156 articles were selected for inclusion. Figure \ref{fig:prisma} displays the PRISMA flow diagram detailing the included and excluded articles at each stage of the review process.\\

Articles needed to be (a) published between the years 2000 and 2023 in English language in a journal, conference proceeding, or book chapter; (b) contain quantitative or qualitative empirical research; (c) measure trust of human subjects in AI. Studies were excluded if (a) they measured trust in something other than AI (e.g., the organization providing the AI); (b) measured other constructs than trust, such as reliance or propensity to trust; (c) did not include actual participants (i.e., a simulated study or proposal); (d) did not adequately report the employed methodology.
After the full-text review, we extracted the following meta-information from the included articles, using the Scopus API: (a) citation information, including author names, document and source titles, and citation counts; (b) bibliographical information, including affiliations and countries; (c) abstract \& keywords, including author keywords; (d) all cited references.


\subsection{Analyses}
\subsubsection{Bibliometric techniques}
 \label{bibliometrics} 

The quantitative techniques used in this work are based on bibliometrics, which uses meta-data, such as number of citations, references, author information, and keywords, to characterize the scientific landscape of a research field \citep{Culnan1986TheID, Nerur2008TheIS}. 
To address \textbf{RQ1}, we analyzed the publishing countries and sources with bibliometric techniques, thereby retrieving the most cited articles and influential works, i.e., those that appear most often within the references of the articles in our dataset.
\\
To address \textbf{RQ2} we used bibliometrics and explored two knowledge structures to gain insights into the scientific landscape of empirical trust in AI research \citep{Liu2014CHI1M, Aria2020MappingTE, Wamba2020AreWP}. We characterized the \textbf{intellectual structure} of the field by analyzing co-citation patterns among articles and sources, thereby determining their degree of relatedness and identifying common schools of thought. We characterized the \textbf{conceptual structure} of the field by analyzing keyword co-occurrences using author keywords, thereby identifying research themes, as well as temporal and thematic trends.

However, we note that bibliometrics does not offer insights into the thematic content of the articles \citep{introbibliometrix}. 
Therefore, to complement the statistical findings from the bibliometric analysis, we conducted a thorough content analysis of the most cited works in our database, which we further describe below.

All statistical analyses were conducted using Python and the R bibliometrix package \citep{Aria2017bibliometrixAR}, while all visualizations were generated using VOSviewer \citep{Eck2009SoftwareSV}.

\subsubsection{Content analysis} To add to the insights from the bibliometric analysis and address \textbf{RQ3}, we conducted a content analysis, in which we systematically coded a sub-set of our core articles regarding specific categories. 
Content analysis is a commonly employed method in tandem with bibliometric methods \citep{Arici_2019, Ridwan_22, Enebechi_2020}. Whereas bibliometrics offers quantitative insights into the literature, such as publication patterns, citation networks, and trends over time, content analysis complements this by allowing to interpret the use of citations within the articles. Together, this mixed-method approach provides a more holistic overview over the research landscape. 
First, building on prior works \citep{FAHIMNIA2015101, Zhao2023LeadingVR, Corallo2019HumanFI}, we selected a representative subset of articles to analyze. To derive the representative subset, we retrieved the ten percent most cited articles for each discipline, ensuring a minimum of five and a maximum of ten entries per discipline. This method accounts for the variations in scholarly output and citation frequencies across different disciplines \citep{CEROVSEK2014147, Nielsen2021}. 
Articles were annotated, using coding categories that have been generated from prior literature reviews and based on the findings from the bibliometric analysis.

\section{Results}\label{results} 

\subsection{Overview}\label{overview}
\begin{figure}[h!]
    \centering
    \includegraphics[width=8cm]{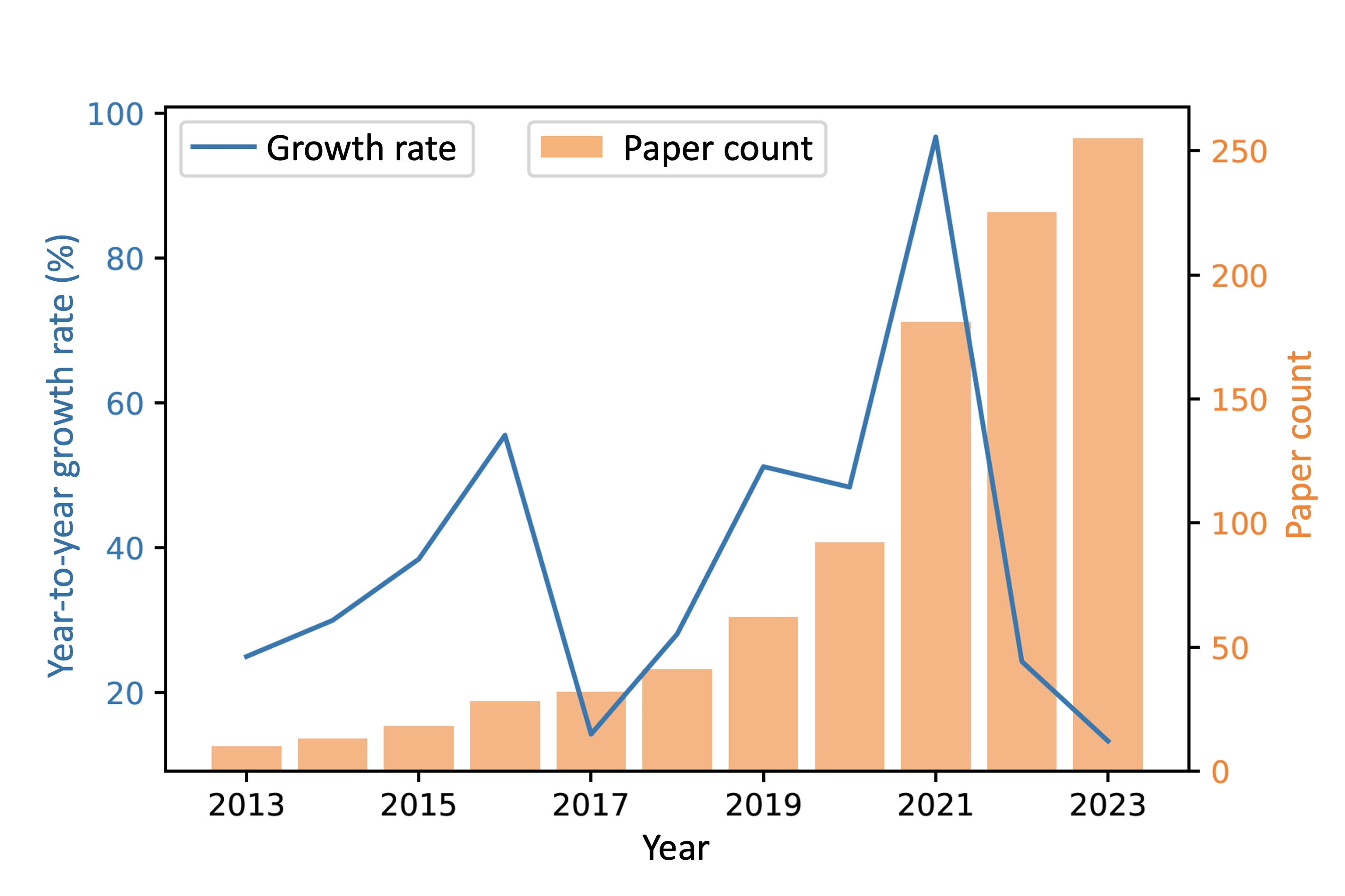}
    \caption{Article count and year-to-year growth rate in the core dataset (2013-2023).}
    \label{fig:articles-growth}
\end{figure}

Our search resulted in a core dataset of 1’156 articles, citing 36’306 articles, which comprise the ``cited dataset''. The core dataset covers a wide range of disciplines, including the physical, social, and life sciences. We show the distribution of research domains in Table \ref{tab:disciplines}. 
The majority of articles (57.8\%) originate in technology-focused research domains, such as human-robot interactions, computer science, or robotics and engineering. Publications in social sciences and psychology contribute to 17.8\% of the total number of retrieved articles. In addition, a few articles originate from transportation and business, management, and marketing. 
In Figure \ref{fig:articles-growth}, we show the number of articles in the core dataset per year and their year-to-year growth rate. In particular, Figure \ref{fig:articles-growth} shows a sustained year-to-year growth in the period 2011-2018, followed by a steep increase in the number of articles focused on empirical research on trust in AI in 2021. However, the growth rate has decreased in 2022 and 2023.

\begin{table}[h!]
    \centering
    \ra{1.3}
    \begin{tabular}{@{}l|l@{}}
       \hline
Research Domain & Articles, n (\%) \\ 
\hline
        Computer Science & 176 (15.2\%) \\
        Human-Computer Interaction & 171 (14.8\%) \\
        Human-Robot Interaction & 148 (12.8\%) \\
        Inter-/Multidisciplinary & 99 (8.6\%) \\
        Robotics & 89 (7.7\%) \\
        Psychology and Cognitive Sciences & 87 (7.5\%) \\
        Healthcare and Medicine & 69 (6.0\%) \\
        Transport and Logistics & 48 (4.2\%) \\
        Information Science & 46 (3.9\%) \\
        Business and Economics & 39 (3.4\%) \\
        Engineering & 34 (2.9\%) \\
        Human Factors \& Ergonomics & 34 (3.0\%) \\
        Behavioral and Decision Science & 29 (2.5\%) \\
        Marketing & 21 (1.8\%) \\
        Environmental and Agricultural Sciences & 15 (1.3\%) \\
        Law, Public Policy and Administration & 8 (0.7\%) \\
        Tourism & 8 (0.7\%) \\
\hline
    \end{tabular}
    \caption{The distribution of the bibliographic records in the core dataset by research domain.}
    \label{tab:disciplines}
\end{table}

\newpage
\subsection{Publication patterns (RQ1)}
\label{pub_patterns}

\subsubsection{Publishing countries}
In Table \ref{tab:countries_top}, we show the distribution of the top 10 publishing countries in the core dataset. We identified the publishing country of each article, by retrieving the country of the respective corresponding author.
Articles from the United States, Germany, and the United Kingdom comprise for slightly more than 56\% of all publications in the core dataset. About 46\% of all articles in the cited dataset originate from those three countries instead. About one third of publications in the core, and nearly one quarter of the cited datasets originate from the remaining countries. Notably, 80.5\% of the articles in the core dataset originate from Global North countries, characterized by the acronym WEIRD (Western, Educated, Industrialized, Rich, Democratic).

\begin{table}[h!]
    \centering
    \ra{1.3}
    \begin{tabular}{@{}l|c|c@{}}
    \hline
    Country & Core articles, n (\%) & Cited articles, n (\%) \\
    \hline
    United States & 405 (34.1) & 11'759 (31.3) \\
    Germany & 150 (12.6)& 2'406 (6.4)\\
    United Kingdom & 114 (10.0)& 2'960 (7.9)\\
    China & 91 (7.7) & 1'422 (3.8)\\
    Netherlands & 60 (5.1)& 1'374 (3.7) \\
    South Korea & 49 (4.1)& 1'316 (3.5)\\
    Japan & 49 (4.1) & 739 (2.0)\\
    Italy & 45 (3.8)& 874 (2.2)\\
    Canada & 44 (3.7) & 1'316 (3.5) \\
    Australia & 36 (3.0)& 1'231 (3.3)\\
    \hline
    \end{tabular}
    \caption{The distribution of bibliographic records by top 10 publishing countries.}
    \label{tab:countries_top}
\end{table}

\subsubsection{Publishing sources}

The majority (56.0\%) of articles in the core dataset are journal publications, followed by conference proceedings (39.1\%) and book chapters (4.9\%). In Table \ref{tab:publishing_source_top10}, we show the top 10 publishing sources in the core and cited datasets, together with their number of citations in the core and cited datasets, as well a their Hirsch-Index \citep{HirschIndex}. About 23\% of these publications are distributed across four conferences (HRI, RO-MAN, CHI, IUI) and a conference proceedings series (Lecture Notes in Computer Science). However, their representation in the cited dataset is minimal. Close inspection reveals a ``long tail'' of 155 distinct sources, illustrating the extent of research dispersion of trust in AI research across diverse application areas. In Figure \ref{fig:bradford} in the Appendix, we display the distribution of these sources, highlighting the ``core'' set of sources in accordance with Bradford's law \citep{Bradford1985SourcesOI, Brookes1969BradfordsLA}. Bradford's law is a bibliometric principles that states that if publishing sources in a field are sorted by their productivity, they can be divided into a productive core, followed by several zones of progressively less productive publishing sources  \citep{Bradford1985SourcesOI}.

\begin{table}[h!]
    \centering
    \ra{1.3}
    \begin{tabular}{@{}p{5cm}lll@{}}
    \toprule
    Source & Core articles, n (\%) & Cited articles, n (\%) & H-Index \\
    \toprule
    ACM/IEEE Int. Conf. on Human-Robot Interaction (HRI) & 63 (5.4) & 371 (1.0) & 25\\
    IEEE Int. Conf. on Robot and Human Interactive Communication (RO-MAN) & 60 (5.2) & 408 (1.1) & 16\\
    Int. Journal of Social Robotics	& 35 (3.0) & 249 (0.7) & 15\\
    ACM Conf. on Human Factors in Computing Systems (CHI) & 32 (2.8) & 520 (1.4) & 15\\
    Int. Journal of Human-Computer Interaction & 31 (2.7) & 125 (0.3) & 9 \\
    Lecture Notes in Computer Science & 31 (2.7) & 582 (1.6) & 8\\
    Computers in Human Behavior	& 25 (2.2) & 453 (1.2) & 16\\
    Int. Conf. on Intelligent User Interfaces (IUI)	& 19 (1.6) & 115 (0.3) & 11\\
    Frontiers in Psychology &15 (1.3) & 158 (0.3) &7 \\
    Transportation Research& 14 (1.2) & 265 (0.7) &7\\
    \hline
    \end{tabular}
    \caption{The distribution of bibliographic records by top 10 publishing sources, sorted by number of core articles.}
    \label{tab:publishing_source_top10}
\end{table}

\subsubsection{Most cited core articles} In Table \ref{tab:most_cited_core}, we show the top 10 most cited articles in the core dataset. The most cited work appears in the premiere conference on knowledge discovery and data mining (ACM SIGKDD) \citep{Ribeiro-et-al_2016} and focuses on machine learning (ML) modelling. In that article, Ribeiro et al. introduce LIME (Local Interpretable Model agnostic Explanations), which is an explanation method for the predictions of ML models. Further, they provide some examples of empirical studies where the effects of LIME on ML model users' trust are analyzed. However, the authors refrain from defining the construct ``trust'', measuring it by asking study participants questions, such as ``would you trust this model prediction?'' or assessing which classifier is chosen, therefore ``trusted'', among different choices \citep{Ribeiro-et-al_2016}. In the second most-cited work in Table \ref{tab:most_cited_core}, Choi and Ji generalize the Technology Acceptance Model including items, such as perceived ease of use and trust, to study users’ adoption of autonomous vehicles \citep{Choi2015InvestigatingTI}. Trust in autonomous vehicles is measured adapting items from existing works in the domains of online shopping and electronic commerce \citep{gefen_psychology_2013,pavlou2003consumer}. As a result, participants to Choi and Ji's study are asked to rate the statements ``Autonomous vehicle is dependable,'' ``Autonomous vehicle is reliable'' and ``Overall, I can trust autonomous vehicle'' on a Likert scale to measure their levels of trust in autonomous vehicles -- see Table 1 in \citep{Choi2015InvestigatingTI}. Similarly, in their work, \citet{Waytz2014TheMI} focus on autonomous vehicles. In a study with a driving simulator, the authors measured participants' self-reported trust in the simulated vehicle using an ad-hoc questionnaire comprising eight items, which included ``how much [the participants] trust the vehicle to drive in heavy and light traffic conditions, how confident they are about the car driving the next course safely'' \citep{Waytz2014TheMI}. In addition, behavioral and  physiological measures, e.g., number of distractions during the simulation and heart rate change, are analyzed in relation to the self-reported trust levels \citep{Waytz2014TheMI}. Interestingly, the remaining most cited works are predominantly recent publications, with three articles focusing on the topic of transparent and explainable AI systems \citep{Shin2021TheEO, Zhang2020EffectOC, Yin2019UnderstandingTE}.

\begin{table}[h!]
    \centering
    \ra{1.3}
    \begin{tabular}{@{}p{5cm}p{2.8cm}p{3cm}l@{}}
    \toprule
Title & Authors & Source & TC  \\
\midrule
``Why Should I Trust You?" Explaining the Predictions of Any Classifier & \citet{Ribeiro-et-al_2016} & ACM SIGKDD & 9'120 
\\
Investigating the Importance of Trust on Adopting an Autonomous Vehicle &
\citet{Choi2015InvestigatingTI} & Int. Journal of Human-Computer Interaction &	700
\\
The mind in the machine: Anthropomorphism increases trust in an autonomous vehicle & \citet{Waytz2014TheMI} & Journal of Experimental Social Psychology	& 693 
\\
The effects of explainability and causability on perception, trust, and acceptance: Implications for explainable AI & \citet{Shin2021TheEO} & Int. Journal of Human Computer Studies & 417 \\

Would You Trust a (Faulty) Robot?: Effects of Error, Task Type and Personality on Human-Robot Cooperation and Trust & \citet{Salem2015WouldYT} & ACM/IEEE Int. Conf. on Human Robot Interaction &  398
\\
Acceptability of artificial intelligence (AI)-led chatbot services in healthcare: A mixed-methods study & \citet{Nadarzynski2019AcceptabilityOA} & Digital Health & 317 
\\
Adoption of AI-based chatbots for hospitality and tourism & \citet{Pillai2020AdoptionOA} & Int. J. of Contemporary Hospitality Mgmt. & 312 
\\
An empirical investigation on consumers’ intentions towards autonomous driving & \citet{Panagiotopoulos2018AnEI} & Transportation Research & 310 
\\
Effect of confidence and explanation on accuracy and trust calibration in AI-assisted decision making & \citet{Zhang-et-al_2020} & ACM Conf. on Fairness, Accountability, and Transparency (FaccT) & 296 
\\
Understanding the Effect of Accuracy on Trust in Machine Learning Models & \citet{Yin2019UnderstandingTE} & ACM CHI & 254 \\
\hline
    \end{tabular}
    \caption{The top 10 most cited articles in the core dataset. TC = Total Count. The number of citations is as of May 2023 (Source: Scopus).}
    \label{tab:most_cited_core}
\end{table}

\begin{landscape}
\begin{table}[h!]
\centering
\ra{1.3}
\begin{tabular}{@{}p{4.5cm}p{2.5cm}p{3cm}p{5cm}l@{}}
  \hline
Title & Authors & Source & Contribution  & TC, n (\%) \\ 
\hline
Trust in automation: Designing for appropriate reliance &  \citet{lee_trust_2004} & Human Factors &  Theoretical model and design principles for achieving appropriate reliance on automated systems & 299 (25.9) \\ 

A meta-analysis of factors affecting trust in human-robot interaction &  \citet{Hancock_2011} & Human Factors &  Survey identifying and synthesizing various factors that influence trust in human-robot interactions & 190 (16.4) \\ 

An integrative model of organizational trust  &  \citet{mayer-et_al_1995} & Academy of Management Review  & Theoretical model of trust within organizations  & 168 (14.5) \\ 

Foundations for an empirically determined scale of trust in automated systems &  \citet{Jian2000FoundationsFA}  & Int. Journal of Cognitive Ergonomics   & Measurement instrument for trust in automated systems & 156 (13.5) \\ 

Trust in automation: Integrating empirical evidence on factors that influence trust & \citet{Hoff2015TrustIA}  & Human Factors  & Survey and model,  integrating empirical evidence on factors that influence trust in automation   & 140 (12.1) \\ 

Perceived Usefulness, Perceived Ease of Use, and User Acceptance of Information Technology & \citet{Davis1989PerceivedUP}  & MIS Quarterly & Technology Acceptance Model (TAM), which explains how perceived usefulness and perceived ease of use influence user acceptance of information technology  & 123 (10.6)\\ 

Humans and Automation: Use, Misuse, Disuse, Abuse & \citet{Parasuraman1997HumansAA} & Human Factors & Review and framework to understand different ways humans interact with automation, including use, misuse, disuse, and abuse.  & 116 (10.0) \\ 

Consumer Acceptance and Use of Information Technology: Extending the Unified Theory of Acceptance and Use of Technology & \citet{Venkatesh2012ConsumerAA} & MIS Quarterly & Unified Theory of Acceptance and Use of Technology (UTAUT), which integrates various models of information technology acceptance into a single framework & 101 (8.7)\\
Measurement Instruments for the Anthropomorphism, Animacy, Likeability, Perceived Intelligence, and Perceived Safety of Robots & \citet{Bartneck2009MeasurementIF} & Int. Journal of Social Robotics & Measurement instruments to assess various attributes of robots, including anthropomorphism, animacy, likeability, perceived intelligence, and perceived safety & 99 (8.6) \\
The role of trust in automation reliance & \citet{Dzindolet2003TheRO} & Int. Journal of Human-Computer Studies & Investigation of the role of trust in users' reliance on automation &  78 (6.7)  \\ 

\hline
\end{tabular}
\caption{The top 10 most influential articles in the cited dataset. TC = Total Count. The percentage indicates the proportion of articles in the core dataset that cite the article.} 
\label{tab:mostcited-articles}
\end{table} 
\end{landscape}

\subsubsection{Influential cited articles}\label{influential}

The results of the analysis of citation trends in the cited dataset reveal ten most influential theoretical contributions, as seen in Table \ref{tab:mostcited-articles}. One quarter of all articles in the core dataset cite the seminal work by \citet{lee_trust_2004}, where a conceptual model of trust in automation is introduced and the concept of trust calibration (correspondence between trust and automation capabilities) is introduced. The authors employ a definition of automation as ``technology that actively selects data, transforms information, makes decisions, or controls processes.'' \cite[p.50]{lee_trust_2004}. While their definition is broad, the authors focus on technologies in work-related settings, with costly or catastrophic consequences when inappropriate or mis-calibrated trust leads to disuse or misuse (e.g., automated navigation systems, autopilots, or aviation management systems). Moreover, highly cited articles include the meta-analysis of factors affecting trust in human-robot interactions by \citet{Hancock_2011} (16.4\%) and the work by \citet{Jian2000FoundationsFA} (13.5\%) that introduces an empirical scale to measure trust in automation. Hancock et al. identify human-, environment-, and robot-related factors that contribute to trust in human-robot interactions \citep{Hancock_2011}, while Jian et al. develop a trust measurement scale by asking university students to rate how certain words relate to (a) trust in general; (b) trust between people; and (c) trust in ``automated systems.'' A cluster analysis revealed 12 factors of trust in automation, which were subsequently used to build the scale. It is important to note, however, that the scale was not validated by the authors \citep{Jian2000FoundationsFA}. Further, Table \ref{tab:mostcited-articles} shows that the majority of influential articles originate in the human factors and ergonomics domain. Most contributions propose theoretical frameworks for understanding trust, followed by comprehensive reviews of factors that influence trust across various contexts, and, finally, empirical studies that quantify and analyze trust using experimental methods. 

An investigation of citation trends over time revealed that the works by \citet{lee_trust_2004} and \citet{Parasuraman1997HumansAA} remain the most cited by the manuscripts in the core dataset across all time periods. Interestingly, the work by \citet{mayer-et_al_1995} was not highly cited before 2012 and became the third most cited one in 2021, indicating a renewed interest in this work. An illustration of the citation trends of the five most influential works can be seen in Figure \ref{fig:top5_overtime} in the Appendix. Additionally, we analyzed the relative importance of these influential articles across different research domains. For instance, the work by \citep{lee_trust_2004} is particularly significant for HCI research, while the work by \citep{Hancock_2011} holds greater importance for HRI. In contrast, for the less represented research domains such as Tourism and Law, and Public Policy and Administration, these works do not appear influential. The results of this analysis are visualized in a heatmap shown in Figure A2 in the Appendix.

\subsection{Intellectual structure (RQ2)}\label{results-intellectual}
To analyze the intellectual structure of empirical trust in AI research, we investigated the co-citation patterns within the cited dataset. We identified two units of analysis: (a) co-cited articles; and (b) co-cited publishing sources. The former allows us to situate influential works in the overall structure of cited articles, while the latter allows us to situate publication outlets among other disciplines. 

\subsubsection{Co-cited articles.}

\begin{figure}[h!]
    \centering
    \includegraphics[width=\textwidth]{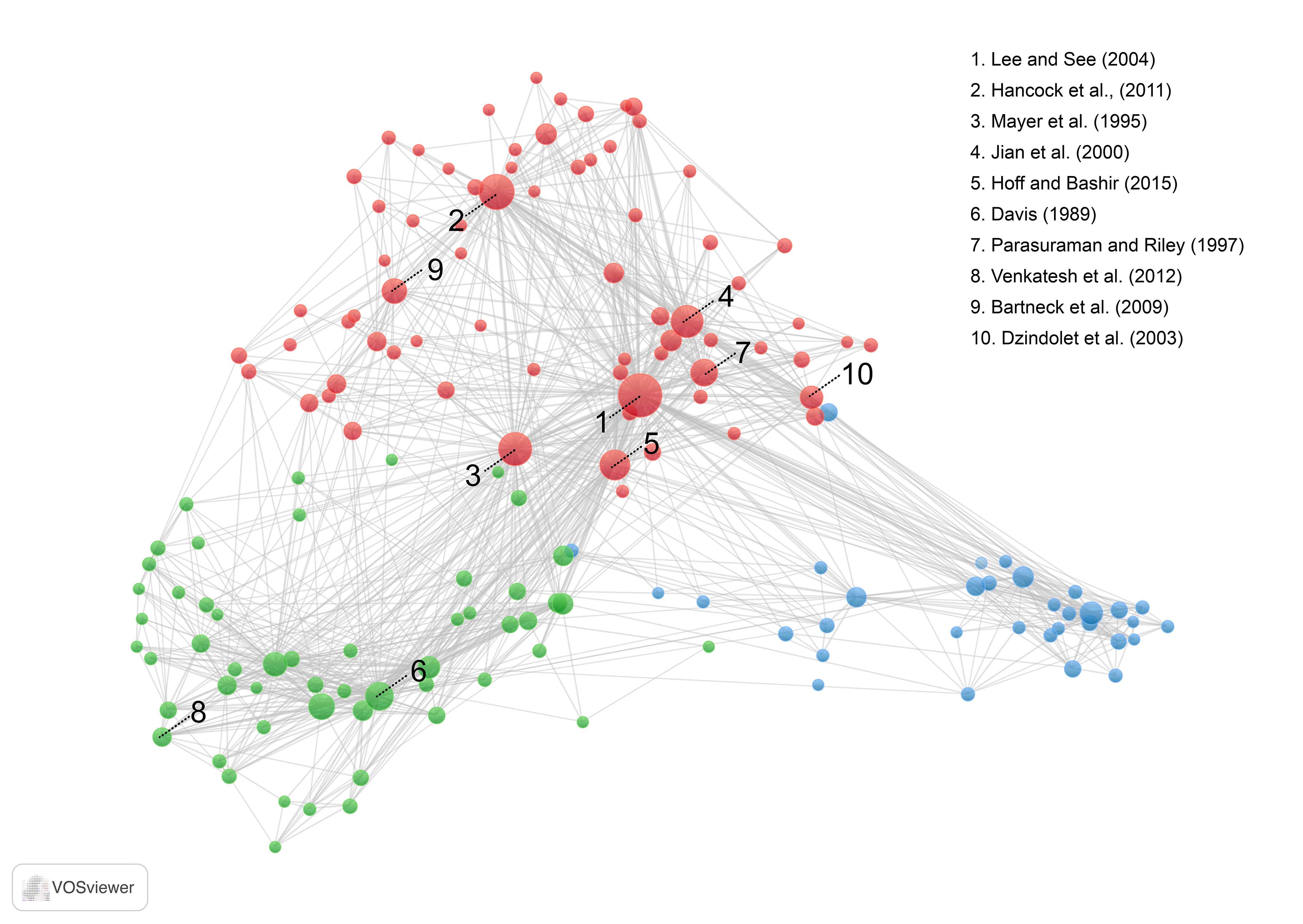}
    \caption{Network visualization of frequently co-cited articles. Each node represents a specific article. Thicker links represent a higher number of co-citations, while larger nodes indicate a greater number of occurrences.}
    \label{fig:co-ciation-network-core}
\end{figure}

In Figure \ref{fig:co-ciation-network-core}, we show the network of articles from the cited dataset for the co-citation analysis. The influential works listed in Table \ref{tab:mostcited-articles} are grouped into two of three primary clusters, with eight out of ten articles belonging to the same cluster.
At the center of Figure \ref{fig:co-ciation-network-core}, the most highly connected and central nodes include the works by \citet{lee_trust_2004}, \citet{mayer-et_al_1995} and \citet{Hoff2015TrustIA}. These studies not only have the highest number of co-citations but also serve as a bridge connecting the different clusters. 
For visibility, we only show the influential works. However, we note that the cluster in blue includes more recent works, including the works by \citet{miller_explanation_2018} and \citet{Ribeiro-et-al_2016}, as well as the work by \citet{dietvorst_overcoming_2018}.

\subsubsection{Co-cited publishing sources.}
\begin{figure}[h!]
    \centering
    \includegraphics[width=\textwidth]{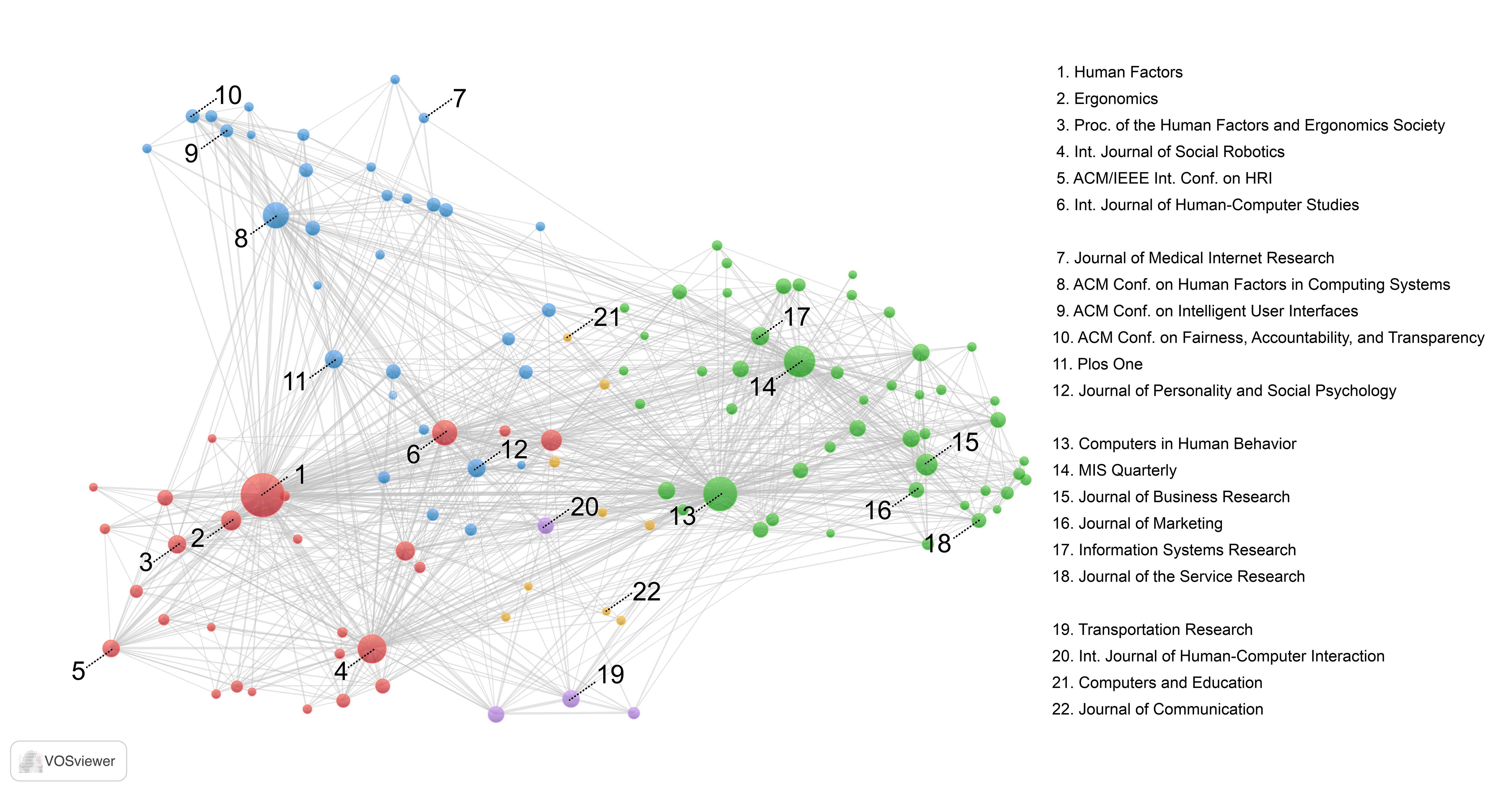}
    \caption{Network visualization of frequently co-cited sources (cited dataset). Each node represents a source in the cited dataset. Thicker links represent a higher number of co-citations, while larger nodes indicate a greater number of occurrences.}
    \label{fig:sources}
\end{figure}

To situate publication outlets of influential works in the overall intellectual structure of cited articles, we analyzed co-citation links between sources (publication outlets), which indicates how the different disciplines of the co-cited works relate to each other. As seen in Figure \ref{fig:sources}, three main clusters emerge. The first cluster (in red, bottom left), called ``human factors and ergonomics,'' sees Human Factors as being the most relevant and connected publishing outlet. 
It is the journal where the influential works \citet{lee_trust_2004,Hancock_2011,Parasuraman1997HumansAA,Hoff2015TrustIA} shown in Table \ref{tab:mostcited-articles} are published. 
Additionally, the cluster includes works from the International Journal of Social Robotics, which are highly co-cited within the network. The second cluster (in green, right), called ``management, technology and economics,'' contains outlets focused on humans interacting with computer systems with application to business, marketing and management. Among them Computers in Human Behavior shows the highest degree of connectivity. In particular, Academy of Management Review belongs to this cluster; the influential work \citet{mayer-et_al_1995} is published there. The final cluster (in blue, top right), called ``Computer Science,'' includes premier conference outlets, such as CHI, IUI, and FaccT. Two smaller clusters appear in the backfround and comprise a few sparse nodes, which include Transportation Research and the Journal of Communication. These outlets appear less frequently and are less connected to others in the network.

\newpage

\begin{figure}[h!]
     \centering
     \begin{subfigure}[b]{10.5cm}
         \centering
         \includegraphics[width=\textwidth]{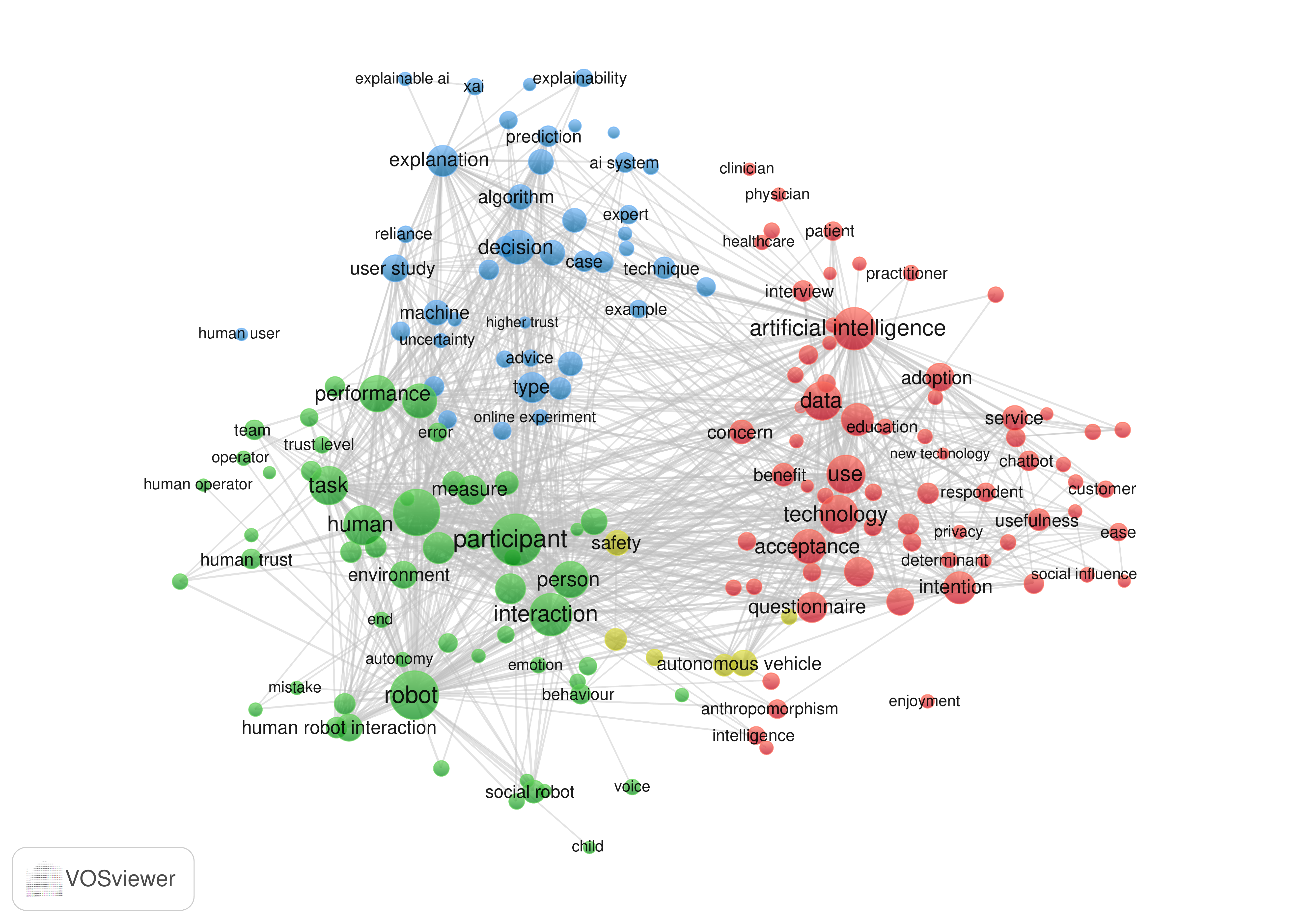}
         \caption{}
         \label{fig:themes-a}
     \end{subfigure}
     \hfill
     \begin{subfigure}[b]{10.5cm}
         \centering
         \includegraphics[width=\textwidth]{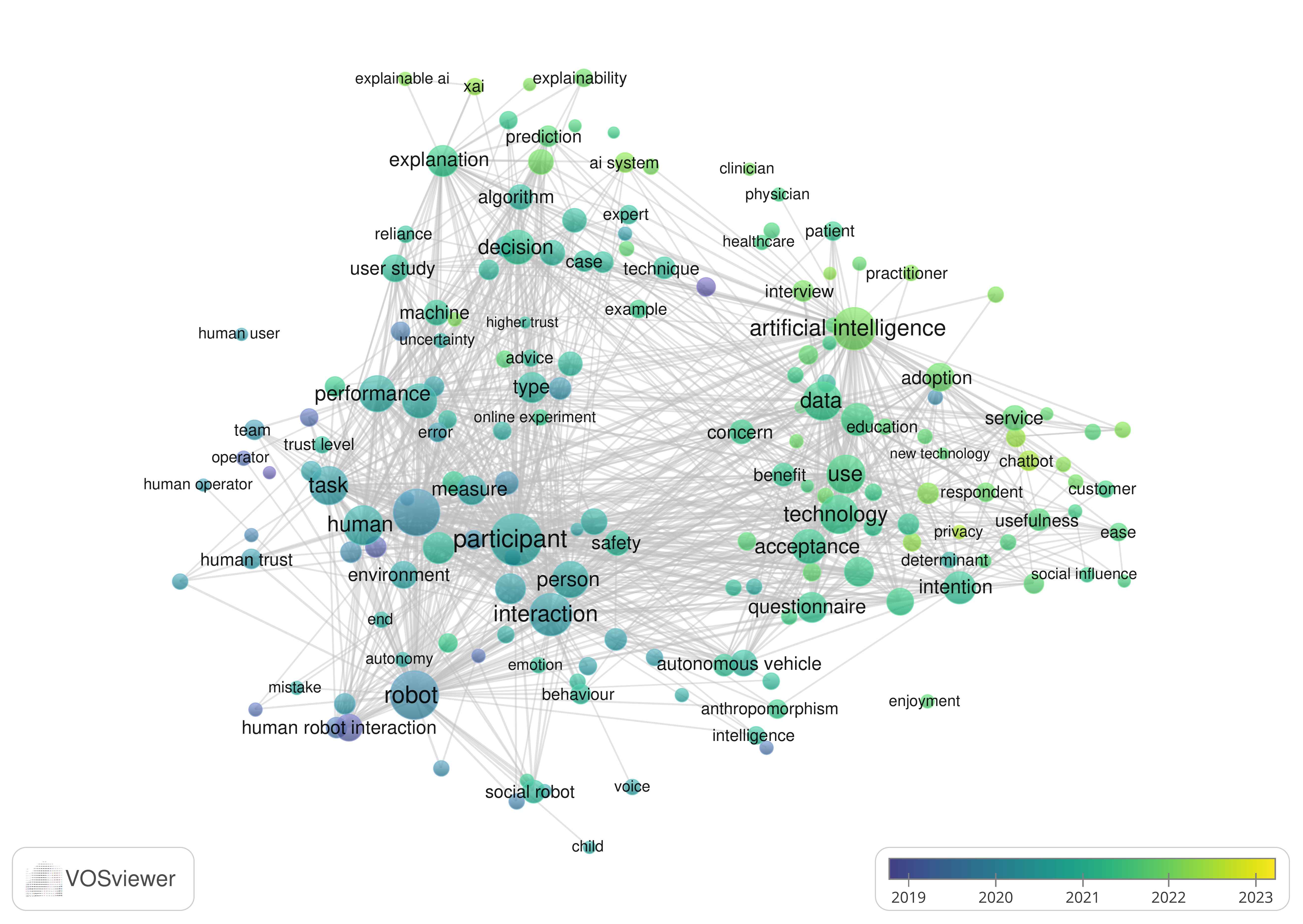}
         \caption{}
         \label{fig:themes-b}
     \end{subfigure}
        \caption{Keyword co-occurrence network based on titles and abstracts. Subfigure (a) shows the co-occurrence and frequency of keywords in the dataset. Subfigure (b) shows the temporal evolution of the keyword co-occurrence network for the years 2019-2023.}
        \label{fig:themes}
\end{figure}
\subsection{Conceptual structure (RQ2)}\label{results-conceptual}

In the following paragraphs, we explore the research themes that emerge from keyword co-occurrences within the core dataset before commenting on recent trends.
Both analyses are displayed in Figure \ref{fig:themes}.

\subsubsection{Research themes}\label{themes}

The keyword co-occurrence network is displayed in subfigure (a) of Figure \ref{fig:themes}. Results of the keyword co-occurrence analysis revealed several topic clusters, which can be summarized into four main research themes (RT). The topics and themes provide an overview of the breadth of empirical research on trust in AI over time.
AI systems that have been investigated over time range from automated fault management \citep{Lorenz2002DisplayIE}, recommender systems \citep{pu2006trust} to robots \citep{Hayashi-et-al_2017} and automated vehicles \citep{Ruijten2018EnhancingTI}. Examples of most recent systems include decision-support tools \citep{Yin2019UnderstandingTE} and explainable AI \citep{Zhang2020EffectOC}. 

For convenience, the researchers assigned labels to each theme. We comment on them in what follows.  \\

\noindent{\textit{RT1: Interactions with robots.}} This theme is broadly focused on interactions with embodied robots in diverse settings, such as service and security robots, e.g., \citep{MeidutKavaliauskien2021TheEO, Babel2021InvestigatingTV, Inbar2019PolitenessCP}, social robots, e.g., \citep{Guggemos2020HumanoidRI, Kim2013CaregivingRI}, or robots in emergency situations \citep{Robinette2016}. For instance, \citet{Inbar2019PolitenessCP} study trust in autonomous  ``peacekeeping'' robots in shopping malls to understand the  willingness of humans to cooperate with them.

Research in this theme also examined different types of embodiments, such as humanoid \citep{Natarajan2020EffectsOA} or non-humanoid robots, including suitcase-shaped robots to support the navigation of users with loss impairment \citet{Guerreiro2019CaBotDA}. 
This theme also includes the highly cited works \citet{Freedy2007MeasurementOT} and \citet{Salem2015WouldYT}. This latter examines how the reliability and personality traits of robotic home companions affect trust in them.\\


\noindent{\textit{RT2: Interactions with multimodal and collaborative systems.}} Research in this theme focuses on interactions with autonomous and multimodal systems, where collaboration, coordination, or teamwork are crucial, such as when operating autonomous vehicles or devices in military operations. AIs in this theme include robotic devices \citep{Nikolaidis2015ImprovedHT}, self-driving cars or other autonomous vehicles \citep{Waytz2014TheMI}, as well as unmanned vehicles \citep{Mercado2016IntelligentAT} and abstract autonomous agents \citep{Dekarske2021HumanTO}. Research in this theme may rely on simulations, given the difficulty of conducting field studies in high-risk settings. For instance, \citet{Freedy2007MeasurementOT} develops a simulation environment to test the performance of mixed human-robot teams in a variety of situations, including military ones \citep{Freedy2007MeasurementOT}. The keyword analysis further highlights the importance of specific human factors that are crucial in shaping successful human-AI collaboration, including appropriate reliance, situation awareness, perceived risk, cognitive load, shared mental models, and information quality or transparency.\\

\noindent{\textit{RT3: AI-assisted decision-making.}} Research in this theme broadly focuses on decision-making under uncertainty with the aid of abstract ``algorithms'' or decision-support systems. They include healthcare or other high-risk applications and focus on what the literature considers as ``modern'' AI systems, e.g., machine-learning models \citep{Yin2019UnderstandingTE}, deep-learning based systems, artificial intelligence (systems)  \citep{Yang2019AttitudesOC}, or decision-support systems. Research in this stream is interested in fostering appropriate levels of trust in AIs, in order to optimize (joint) performance. For instance, \citet{Yin2019UnderstandingTE} examine how a model's stated performance affects people's understanding and trust in it. 
This theme includes the highly-cited work \citet{Ribeiro-et-al_2016}. Keywords in this theme include ``explainable AI'' \citep{Zhang2020EffectOC, Kenny2021ExplainingBC, Yang2020HowDV} ``interpretability'' \citep{kaur-et-al_2020}, as well as ``trust calibration'' \citep{Yang2020HowDV}.\\

\noindent{\textit{RT4: AI in consumer settings.}} This theme includes research on the use of AIs in the consumer context, which includes recommender systems  \citep{Pu2007TrustinspiringEI, Zanker2012TheIO}, intelligent tutoring systems \citep{Wang2020ParticipantOS}, chatbots \citep{Nadarzynski2019AcceptabilityOA}, and other conversational agents \citep{Hoegen2019AnEC}. 
Research explores how various design factors may affect trust and other human factors, such as usability, user experience, or satisfaction. Desired outcomes include increased technology acceptance or adoption \citep{Hegner2019InAW}, or purchase intentions. For instance, \citet{Zanker2012TheIO} conducted an online experiment to study the effect of the provision of explanations on trust in and intention to use a recommender system for spa resorts. Here, the authors operationalise the construct trust in AI as trust in information provision by the recommender system (``I trust in the information that I receive from the [recommender system]'') and worry about data usage (``I do not worry about the use of my input data to the [recommender system]''). 

\subsubsection{Recent trends}\label{recent_trends}

Recent trends are visualized in the keyword co-occurrence network in subfigure (b) of \ref{fig:themes-b}.
An analysis of the development of the conceptual structure over the last five years in our dataset (2019-2023) reveals a number of additional trends. First, the visualization reveals that numerous keywords have emerged within the past two years alone, signifying a phase of rapid development in AI research. Reflected in these developments are, among others, the impact of the COVID-19 pandemic and advancements in novel technologies like large language models, mixed reality, and voice assistants. Additionally, there's a growing emphasis on critical themes such as explainability, transparency, and fairness in AI. For example, \citet{Laxar-et-al_2023} explore the influence of explainable clinical decision support systems on physicians' trust in decision support systems in the context of rapid triage decisions, such as those that impacted Covid-19. Similarly, \citet{MeidutKavaliauskien2021TheEO} examine how fear of COVID-19 affects trust and intention to use airport service robots. Moreover, \citet{Wu2023HowII} explore students' trust and usage intentions in the context of interactions with ChatGPT.
Overall, topics related to RT1, such as human-robot interaction, as well as recommender systems in RT4 have an older average publication year, while topics related to RT3 have a more recent average publication year.

\subsection{Content analysis (RQ3)}
\label{results-qual}
In the following, we report our findings from the content analysis of the 128 most cited articles in our core dataset. We split them into elements pertaining to (a) theoretical foundation and (b) research design.

\subsubsection{Theoretical foundation}

In nearly half of the articles (N=56, 43.8\%) no theoretical model of trust in AI is applied. That is, in these works, the authors measure ``trust in AI'', but do not formulate hypotheses to motivate their empirical studies and do not relate their findings to any established theory on trust in human-AI interactions. 30 of these articles originate in technical research domains, such as HCI, HRI, Computer Science, while the rest is distributed across non-technical domains, such as Marketing, Health, and Law, Public Policy and Administration. Often, these works explore whether or to what degree a change in trust can be observed when interacting with AI system featuring different design elements \citep{kaur-et-al_2020, Yang2017, Cai2019, Ribeiro-et-al_2016, Salem2015WouldYT}.

Of the remaining 73 articles, 38 (30.0\%) extend a theoretical model to include trust in technology or a system. Specifically, the Technology Acceptance Model or TAM is used by 13 studies \citep{pavlou2003consumer} and the Unified Theory of Acceptance (UTAUT) \citep{UTAUT} is used by 10 studies. The remaining articles formulate hypotheses based on theories that are not directly related to trust, such as the the Computers are Social Actors (CASA) theory \citet{Nass1994ComputersAS}, used by 5 studies, or based on prior empirical works. \\
Only six articles explicitly formulate hypotheses about trust in AI that are grounded in theoretical models of trust. For example, \citet{Zhang-et-al_2020} build on theoretical works on trust in automation, including that of \citet{lee_trust_2004}. Based on these works, the authors posit that algorithmic confidence information and explanations would have a positive effect on trust calibration in the context of AI. The authors use behavioral measure of trust (i.e., participants' agreement and switch percentage when presented with algorithmic advice), to model \citet{lee_trust_2004}'s concept of trust calibration. Their findings indicate that explanations can improve trust calibration. However, in their discussion section, they do not explicitly relate their findings back to theoretical considerations by \citet{lee_trust_2004}. 

Additionally, some works justify measuring trust solely by referring to empirical trends. For instance \citet[p.131]{Choi2011TheIO} state ``[t]rust, perceived usefulness, and satisfaction are the most widely used constructs for measuring user evaluations.'' Similarly, \citet[p.131]{Rau2009EffectsOC} state ``[Situation  awareness,  self-confidence,  subjective  workload,  and  trust] are typically mentioned in the literature as critical for assessing human-robot team performance.''
Finally, we note that, while the theoretical underpinnings of trust are frequently introduced in the opening sections of the 128 most cited articles in the core dataset and trust is subsequently measured, an in-depth discussion on how these findings improve our theoretical understanding of trust is frequently lacking in articles' discussion and conclusion sections.

\subsubsection{Research design} 
The vast majority of the 128 most cited articles include a single study (N=108, 84.4\%) with a one-time measurement of trust in AI (N=88, 68.6\%). Exceptions include the work by \citet{Zhang-et-al_2021_Marketing} who conducted three experiments to compare perceptions of human vs. robo-advisors in the context of financial services, or the work by \citet{Buccinca2020ProxyTA}, who conducted three experiments to study the misleading nature of proxy tasks in evaluating explainable AI systems.

Furthermore, quite a few studies develop their own questionnaire items (N=35, 27.3\%). However, the majority of studies (N=74, 57.8\%) use standardized questionnaires. Among these, the questionnaire by \citet{Jian2000FoundationsFA}, see Table \ref{tab:mostcited-articles}, is the most used (N=17). The rest of the studies show a diverse range of measurement sources (51 unique sources), with each source typically cited by no more than one or two studies.

Lastly, trust is frequently measured among several variables, with human performance (N=33, 25.8\%) being the most common. A performance-related keyword measured in four studies is compliance. Other variables describe (a) attitudes toward the AI, e.g., ``perceived usefulness'' (N=16, 12.5\%), and ``perceived ease of use'' (N=14, 11.0\%); (b) personal or situational factors, e.g., perceived risk (N=11, 8.6\%), cognitive load (N=5, 4.0\%), and (c) perceptions of various AI capabilities, including transparency and explainability (N=10, 7.8\%), and ``intelligence'' (N=9, 7.0\%).

\vspace{0.5cm}

\section{Discussion: The Elephant(s) in the Room}
\label{discussion}

In the following, we identify key trends from our analysis of empirical research on trust in AI. These trends highlight critical, often overlooked issues, or ``elephants in the room,'' that researchers, practitioners, and policymakers need to address. We formulate these issues into question. In Section \ref{strategies} we propose strategies for their resolution in a subsequent section.


\subsection{\textbf{Publication patterns: Is trust in AI WEIRD and techno-centric?}}
Our descriptive analysis of the publication patterns reveals a dynamically evolving research landscape, characterized by a steady increase in publication volume, which indicates a growing interest in trust in AI. The research production appears to have peaked in 2021, with the subsequent decline in growth rate suggesting possible stagnation. The research production has primarily been driven by a Western- and global North-led discourse, otherwise described by the acronym WEIRD, which mirrors a broader issue known as the ``AI divide'' \citep{World-Economic-Forum}, wherein a few countries lead the research and development of AI. Moreover, the distributions of bibliographic records and most cited articles in the core dataset show that empirical research on trust in AI is mostly driven by technical domains, e.g., computer science. This trend is amplified by the growing emphasis on explainable AI. This research in turn primarily builds on seminal articles from trust in automation research, as shown by the analysis of the influential articles in the cited dataset \citep{lee_trust_2004,Hancock_2011,Jian2000FoundationsFA,Parasuraman1997HumansAA}.

These findings prompt several important questions. First, what values characterize WEIRD perspectives and are therefore embedded in the empirical research? For instance, do studies assume that technological advancement is inherently beneficial and, thus, follow a strategic agenda \citep{Laux2023TrustworthyAI}? Second, what perspectives are overlooked? Are we missing insights and driving factors of trust from non-Western cultures or marginalized communities? Third, what techno-centric methodologies and assumptions underpin the empirical research? For instance, is trust merely being utilized as an additional performance metric of AI for optimization? \\
Not addressing these questions will likely perpetuate the dominance of WEIRD perspectives, promoting values and and relying on assumptions that may not be globally applicable. This can exacerbate the AI divide, leading to policies and technologies that fail to consider the needs of non-Western communities. Valuable insights from these perspectives will be missed, hindering the development of globally effective trust frameworks. The research’s current techno-centric focus risks reducing trust to a mere performance metric, overlooking broader social, ethical, and humanistic aspects. Additionally, the observed peak in research production in 2021, followed by a decline in the year-to-year growth rate, suggests potential stagnation. Without integrating diverse perspectives, the field may struggle to innovate and address emerging challenges, limiting its evolution and societal impact.

\subsection{\textbf{\textbf{Knowledge structures: Is trust in AI a moving target?}}}

The intellectual structure of empirical research on trust in AI is shaped by three main disciplinary perspectives: human factors and ergonomics, social robotics, and management, technology and economics, with the former being the most prominent. Despite their distinct viewpoints, e.g., trust between humans in organizations \citep{mayer-et_al_1995} versus trust of human operators in automation \citep{Parasuraman1997HumansAA}, the frequent co-citation and centrality of a few influential works indicate their substantial impact on the field.

Additionally, the research themes emerging from the conceptual structure analysis underscore the diversification of application areas for AI. Investigating trust in AI therefore represents a continuously evolving challenge within empirical research, largely because technological advancements are constantly shifting the landscape. On the one hand, these findings indicate a collective interest in consolidating an overarching understanding of trust in AI and a continuous interdisciplinary relevance of the seminal articles. On the other hand, these influential works often serve as foundational references rather than to inform and contextualizing the study. For examples, the influential works are leveraged to inform the definition of trust in the introductory section of an article but are not discussed further, suggesting a lack of deeper theoretical integration to advance the field.

Collectively, this prompts the question whether trust in AI may be a ``moving target'' where researchers feel pressured to keep up with continuous technological advancements and spend insufficient time exploring any single area of trust in AI and its connection to the theoretical foundations in depth. Indeed, novel technologies, such as generative AI, may exacerbate the challenge posed by the rapid pace of AI development, which often surpasses the pace of scientific publication. Moreover, the contexts in which AI technologies operate differ significantly from traditional contexts typically seen in trust-related automation research or interpersonal settings. Historically, the former traditional context is characterized by automating and facilitating prolonged human tasks in work settings \citep{Chignell2022TheEO, Hoff2015TrustIA}. Here, relational dynamics such as the articulation of intent or alignment of goals were often secondary, as human operators retained control over well-defined tasks \citep{Chignell2022TheEO, Chiou2021TrustingAD}. In contrast, the emerging technology landscape shows the growing importance of AI systems that assume various roles within dynamic socio-technical contexts, such as explainable AI \citep{Papagni2022ArtificialAE}. 
It remains an open question if and to what extent the influential works are applicable to the broad spectrum of human-AI interactions.
Not addressing this gap could result in a limited and fragmented understanding of trust across different disciplinary perspectives, such as human factors and ergonomics, social robotics, and management, technology, and economics. While influential works have substantially impacted the field, their use often remains superficial, serving as foundational references without deeper theoretical integration. This can hinder the development of a cohesive framework for understanding trust in AI. As AI technologies rapidly evolve, researchers may feel pressured to keep up with advancements, leading to insufficient exploration of any single area and its theoretical underpinnings. This challenge is exacerbated by novel technologies like generative AI, which outpace scientific publication rates. Without addressing these issues, the field risks stagnation and a lack of progress in effectively understanding and fostering trust in AI.

\subsection{\textbf{Qualitative content analysis: Is ``trust'' simply a convenient word?}} Our content analysis additionally reveals a gap between the theoretical frameworks introduced by influential works and their use in empirical works. In fact, despite the widespread (co-)citation of the influential works, the vast majority of articles in the core dataset do not formulate a theoretical model or hypotheses on trust in AI based on these works, and a large proportion do not introduce any model of trust. Moreover, trust is often measured along other variables, most commonly performance, but without connecting them theoretically. Additionally, results on trust are often only briefly discussed, leaving readers without a deeper exploration of the concept and without a clear perspective on the reproducibility of results or their generalizability.

Collectively, these findings suggest that empirical research on trust in AI has primarily adopted an exploratory approach that still persists after more than two decades of active contributions. Exploratory research, while highly valuable for research to flourish, usually serves as a preliminary, less expensive, at times less reproducible approach aimed to point toward promising directions, which can then be exploited in extensive studies \citep{Shiffrin2017ScientificPD}. It is expected to mature and either inform translational studies that turn discoveries into applications, or to inform development of theoretical frameworks \citep{Shiffrin2017ScientificPD}. However, a large proportion of empirical works on trust in AI contain only a single study to evaluate trust, replicating the finding by  \citet{Vereschak2021HowTE}. Additionally, trust is often operationalized using a diverse array of measures, many of which are adapted from other research fields or created as ad-hoc single-item questionnaires. Indeed, the current state of research prompts the question whether ``trust'' is not simply being used as a convenient word--a concern previously raised by \citet{Hoffman2017ATO}--and one which does not predict whether or not people will use AI or appropriately rely on it, as highlighted by \citet{Kreps2023ExploringTA}.
Without addressing these issues, trust in AI may continue to be used as a convenient but conceptually shallow term, failing to provide meaningful insights into user behavior and the effective deployment of AI technologies. Prominent efforts, such as calibrating trust in the trustworthiness of AI \citep{lee_trust_2004,Jacovi2020FormalizingTI}, would remain affected by inconsistently measured outcomes.

\section{Strategies for Future Work} \label{strategies}

\subsection{Considering missing viewpoints and hidden assumptions}

Understanding trust in AI from diverse perspectives is crucial, especially for effective global policies, particularly as AI performance continues to advance and integrate into various critical societal applications \citep{Feher2024ModelingAT}. In fact, a recent study by the University of Queensland and KPMG found significant trust disparities in AI among emerging economies like Brazil and India, compared to Western countries \citep{Gillespie2023TrustIA}. Cultural background influences trust \citep{Rau_2009}, but these variations are not well understood. To address this, we recommend fostering inclusive discussions, supporting partnerships with underrepresented regions, and developing and applying uniform measurement tools.\footnote{Note, for example, that the German translation of the questionnaire by \citet{Jian2000FoundationsFA} recommends dividing trust and distrust into separate factors \citep{Phler2016ItemanalyseUF}, in contrast to the original English version, which considers them as opposites of the same factor \citep{Jian2000FoundationsFA, Spain2008TowardsAE}, hindering comparability.} Here, we endorse \citet{Linxen2021HowWI}'s recommendations for varied participant samples and cross-cultural study replication. Additionally, we recommend that technical fields actively engage in critically examining the assumptions driving their research production.

\subsection{Contextualizing trust in AI} 
A key challenge is that a one-size-fits-all solution for integrating theoretical models of trust with diverse, evolving AI technologies, is not feasible, as evidenced by the ongoing lack of tailored approaches in current research. Contextualizing theoretical models to different AI systems is crucial for developing an in-depth understanding of trust in AI and addressing unexpected findings \citep{Hong2014AFA}, especially given recent inconclusive results on trust in the domain of explainable AI research \citep{Langer2022}. We recommend avoiding "generic overviews" on trust in AI and instead focusing on specific theoretical contributions that inform empirical studies. This involves:

\begin{enumerate}
    \item developing and integrating context-specific models of trust, tailored to the distinctive characteristics of various AI systems and dynamics of human-AI interactions;
    \item assessing the applicability of these models and analyzing the outcomes of their empirical testing across various interactions and domains.
\end{enumerate}

To provide more concrete guidelines for context specific theorizing specific to the context of trust in AI, we have adapted the recommendations outlined by \citep{Hong2014AFA} to fit this particular domain. These adapted guidelines are thoroughly detailed in Table \ref{tab:guidelines-theory} in the Appendix.

\subsection{Shifting from exploratory to explanatory research}

In addition to the assertion that the absence of theory and robust data leaves AI researchers unprepared to test specific hypotheses in novel contexts \citep{Leichtmann2022EffectsOE}, our findings indicate a different challenge: the dynamic nature of AI itself complicates the development of a theory of trust in AI. This is further perpetuated by newer developments in generative AI. 
We argue that a strategic shift from primarily exploratory research to more robust and explanatory studies is necessary for the field to mature. To do so, empirical trust in AI research can learn from other domains that have faced similar challenges, such as research in Information Systems \citep{Dub2003RigorII, Djamasbi2018BridgingAB}. Possible strategies include (a) the formulation of clear (``why''-)questions on trust that delve into explanatory aspects; (b) the promotion of multiple-study design \citep{Vereschak2021HowTE}, and (c) use of single studies to challenge existing assumptions instead. Furthermore, the consistent implementation of standardized methodologies and best practices is essential for harmonizing research efforts in this area.

We have summarized the analyses emerging from our work, the elephants in the room, and our strategies for future empirical research on trust in AI for the interested reader in  Table \ref{tab:summary-table}.


\section{Limitations}
\label{limitations}

While this study aimed to provide a comprehensive overview over two decades of empirical trust in AI research, we wish to acknowledge a number of limitations of this work. 
First, although we made considerable efforts to ensure an exhaustive literature search on empirical trust in AI research, it is still possible that our analysis may have been limited by researcher's biases or oversights. For instance, our search criteria may have inadvertently excluded relevant AI technologies or disciplinary perspectives. Second, the focus on mapping the research landscape through quantitative measures limited this study's ability to explore certain nuances. For example, it did not consider the potential influence of different trustor types (e.g., user vs. patient) that may be characteristic of various disciplinary approaches. Additionally, it did not examine different causal directions in experimental studies, such as trust being treated as a dependent or independent variable. Third, our full analysis took into account research until and including 2023 and may not account for most recent developments on trust in AI. 
Nevertheless, recognizing that the breadth of novel technologies is continuously evolving, we believe our analysis offers a timely perspective of trends and patterns that can inform future research. We invite the research community to build upon and extend our findings. 

\section{Conclusion}
\label{conclusion}

The growing interest in trust in AI has led to the establishment of an empirical research community focused on the study of this topic. To inform the design of safe and reliable AI that people may trust, it becomes increasingly important that the research community synchronize their efforts and align on the path ahead. This work aimed to contribute to this endeavour by conducting a bibliometric and content analysis of over two decades of empirical research on trust in AI. Based on a thorough mapping of the research landscape, revealing citation patterns, trends and their implications for the research community, we highlight a number of trends and unsolved issues that we named `elephants in the room.' 
Future research should prioritize the in-depth understanding of the construct ‘trust in AI’ through the development of inclusive, methodologically robust, and contextualized theoretical frameworks of trust in human-AI interactions. In fact, the consistent use of theoretical models and robust methodologies, such as clear explanatory questions on trust and multiple-study designs, is essential for harmonizing research efforts and increasing the maturity of the trust in the AI field. Further it will also promote theoretically-sound and evidence-based policies for a responsible use of this technology.
We hope that this work contributes to furthering research on trust in AI and paves the way forward toward the development and design of AI systems people may use and safely interact with. 

\section{Data availability statement}

The datasets used for the analyses are available on OSF: \url{https://osf.io/5k4rf/?view_only=2daf814f14b44c9fa6ef26ee2d0b5144}. 

\begin{landscape}
\newcolumntype{L}{>{%
       \RaggedRight\everypar{\hangafter=1}}X}
    \begin{table}[htpb!]
    \setlength\tabcolsep{2mm}
    \begin{tabularx}{\linewidth}{@{} p{0.21\linewidth}LL @{}}
    \toprule
\textbf{Analysis } & \textbf{Finding} & \textbf{Strategy} \\
    \midrule
    \textit{Publication Patterns (RQ1)}\par \vspace{1mm}
    \hspace{3mm}\textbullet Publishing countries\par \vspace{1mm} 
    \hspace{3mm}\textbullet Publishing sources\par \vspace{1mm} 
   \hspace{3mm}\textbullet Most cited core articles \par \vspace{1mm} 
    \hspace{3mm}\textbullet Influential cited works \par \vspace{1mm} 
    &  \textbf{Is Trust in AI WEIRD and techno-centric?}\par \vspace{1mm}
       Our descriptive analysis of publication patterns reveals a dynamically evolving research landscape with a steady increase in publication volume, peaking in 2021, and driven primarily by a Western- and global North-led discourse, reflecting the broader "AI divide" issue. The empirical research on trust in AI is driven by technical domains like computer science. This raises critical questions about embedded values, overlooked perspectives, and the techno-centric methodologies underpinning this research. \par \vspace{1mm} &
       \textbf{Considering missing viewpoints and hidden assumptions}\par \vspace{1mm}
    Encourage inclusive discussions through partnerships with institutions from underrepresented regions; promote standardization of measurement tools to allow for comparative studies; promote varied participant samples and replication studies as suggested by \citet{Linxen2021HowWI}. \\
    \midrule
   \textit{Knowledge structures (RQ2) }\par \vspace{1mm}
   \hspace{3mm}\textbullet Co-cited articles\par \vspace{2mm} 
\hspace{3mm}\textbullet Co-cited publishing sources\par \vspace{2mm} 
\hspace{3mm}\textbullet Research themes\par \vspace{2mm} 
\hspace{3mm}\textbullet Research trends\par \vspace{2mm} 
   
   &  \textbf{ Is trust in AI a moving target?}\par \vspace{2mm}
   The intellectual structure of empirical research on trust in AI is shaped by main disciplinary perspectives: human factors and ergonomics, social robotics, and management, with human factors being the most prominent. Additionally, a fourth perspective, Explainable AI, is emerging as a significant area of influence. Despite the evolving nature of trust in AI driven by technological advancements, frequent co-citation of influential works indicates a need for deeper theoretical integration to keep pace with rapid AI developments and diverse application areas.&
   \textbf{Contextualizing trust in AI}\par \vspace{1mm}
   Refrain from providing generic overviews on trust in AI. Instead, discuss and integrate specific theoretical contributions that directly inform the empirical study at hand.\par \vspace{2mm}  
    Develop context-specific models of trust tailored to different AI systems and human-AI interactions. Assess and empirically testing these models across various interactions and domains.

 \\
    \midrule
    \textit{Content analysis (RQ3)}\par \vspace{1mm} 
    \hspace{3mm}\textbullet Theoretical foundation\par \vspace{2mm}
    \hspace{3mm}\textbullet Research design \par \vspace{2mm}
    &\textbf{Is ``trust'' simply a convenient word?}\par \vspace{2mm} 
    Our content analysis reveals a significant gap between theoretical frameworks and their application in empirical works on trust in AI, with many studies failing to develop theoretical models or hypotheses. Despite widespread citation of influential works, most empirical research adopts an exploratory approach without deeper theoretical connections, often using ad-hoc methods and providing limited discussion on trust, raising concerns about the term's usage and the reproducibility and generalizability of findings.
    &
    \textbf{Shifting from exploratory to explanatory research}\par \vspace{1mm}
    Transition to confirmatory research and 'why' questions, enabling researchers to uncover the underlying mechanisms and factors that influence trust in AI systems; establish methodological standards and best practices.
    
    \\

    \bottomrule
    \end{tabularx}

    \caption{Summary of analyses, ``elephants in the room'', and strategies for future empirical research on trust in AI. ``Elephants'' and strategies are marked in bold.}
    \label{tab:summary-table}
    \end{table}
    \end{landscape}

\newpage
\appendix
\section*{Appendix}

\setcounter{table}{0}
\setcounter{figure}{0}

\renewcommand{\thetable}{A\arabic{table}}
\renewcommand{\thefigure}{A\arabic{figure}}

\begin{figure}[h!]
    \centering
    \includegraphics[width=\textwidth]{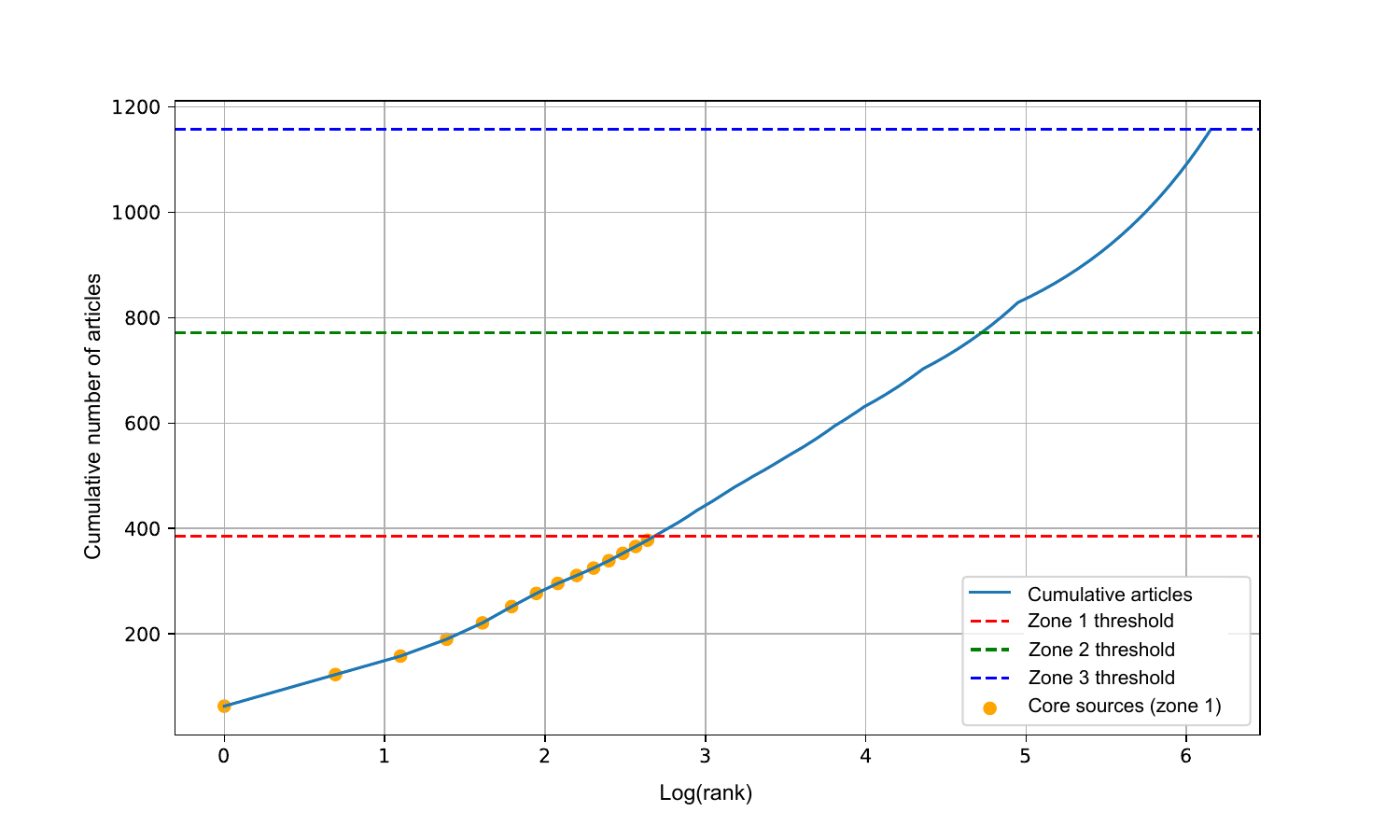}
    \caption{Cumulative number of articles per source. Core sources are identified by using Bradford's Law \citep{Bradford1985SourcesOI}.}
    \label{fig:bradford}
\end{figure}

\begin{figure}[h!]
    \centering
    \includegraphics[width=\textwidth]{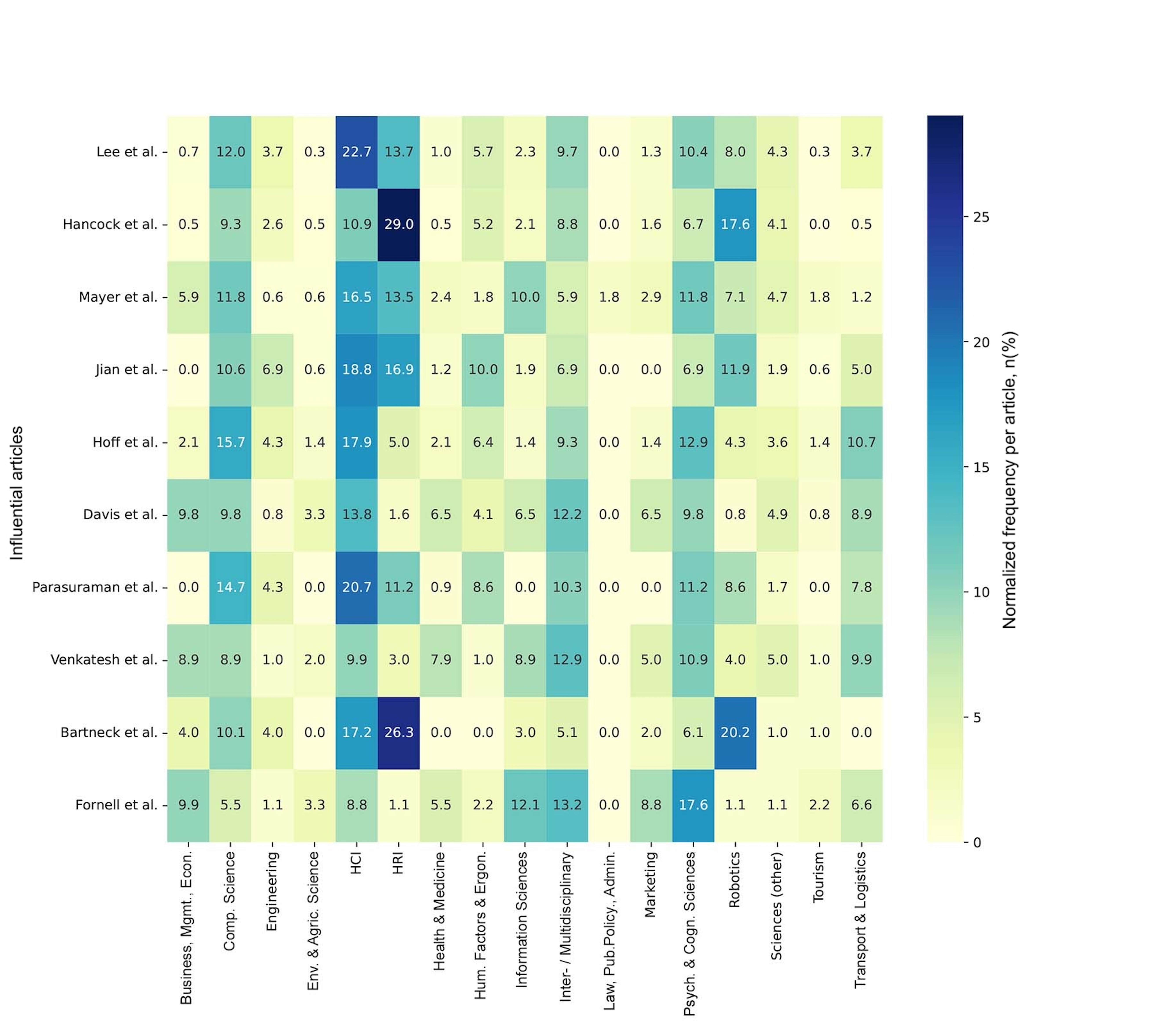}
    \caption{Heatmap, showing the relative importance of the influential works in Table \ref{tab:mostcited-articles} per research domain.}
    \label{fig:heatmap}
\end{figure}

\begin{figure}[h!]
    \centering
    \includegraphics[width=\textwidth]{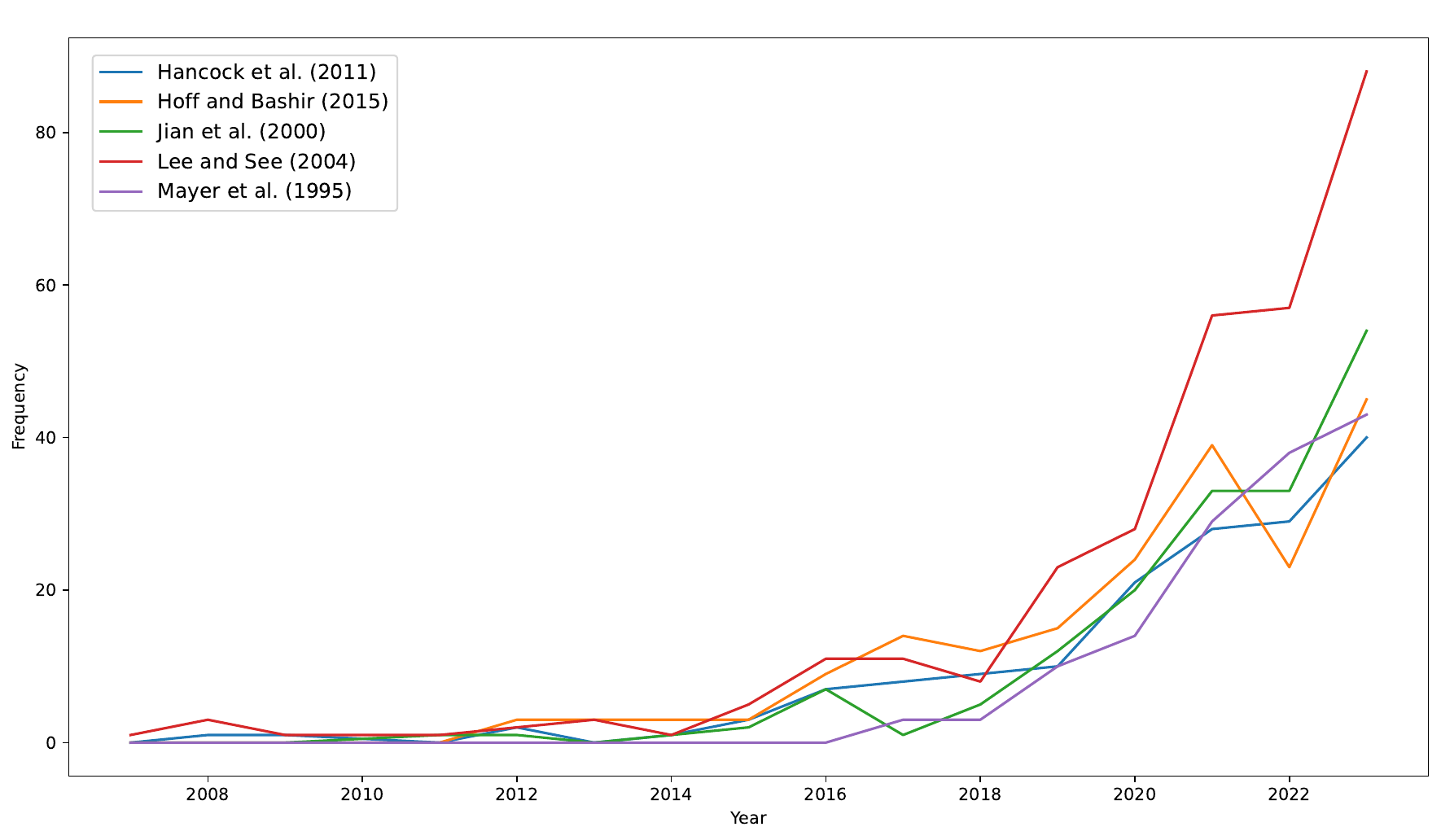}
    \caption{Citation trend of the five most influential papers over time.}
    \label{fig:top5_overtime}
\end{figure}

\begin{table}[]
    \centering
    \ra{1.3}
    \begin{tabular}{@{}p{3.5cm}p{9cm}@{}}
    \toprule
    \textbf{Guideline} & \textbf{Description}\\
    \toprule
    \multirow{2}{3.5cm}{(1) Selection of a general theory of trust} &  \textit{Which theory fits the context best?}\\\cmidrule{2-2} 
     & Leveraging a foundational theory that is relevant for the use-case of interest. Our analysis reveals that the works by \citet{lee_trust_2004}, \citet{mayer-et_al_1995} and \citet{Parasuraman1997HumansAA} are considered foundational. Emerging models, such as \citep{Chiou2021TrustingAD, ferrario_ai_2019, loi2023much} could also be considered. Broader theories of human behavior may be scrutinized for their relationship to trust.\\
     \midrule
     \multirow{2}{3.5cm}{(2) Contextualizing and refining a general theory of trust} & 
     
    \textit{Which elements of the theory fit the context, which do not?}\\\cmidrule{2-2} 
     
     &Ensuring that the theory includes the core variables that are critical to the targeted context, by  exploring   its   conditions   or relaxing  its  assumptions. Existing research that has already explored the boundaries of a given theory for different contexts can inform this step. For example, \citet{Jacovi2020FormalizingTI} explore the role of risk in establishing trust within the context of explainable AI. \\
    \midrule 
    
    \multirow{2}{3.5cm}{(3) Thorough evaluation of the context to identify context-specific factors that should be integrated} & 
     
    \textit{Which factors must be integrated into the theory for it to correspond to the context?}\\\cmidrule{2-2} 
     
    &Identify key factors specific to the context. Drawing on our analysis in Section \ref{results-conceptual} and recent studies, crucial factors include (a) the human user (e.g., user expectations, knowledge, and interaction role); (b) the AI system (e,g., type, functionalities, embodiment \citep{Glikson2020HumanTI}, automation level \citep{simmler2021taxonomy}; (c) the interaction's relational dynamics (e.g., goals, success criteria, trustworthiness criteria of the AI\citep{Salimzadeh_2023}, socio-technical context \citep{Ehsan2020HumancenteredEA, benk2022value}, and finally (d) the risk assessment (e.g., interaction risks \citep{Jacovi2020FormalizingTI}, possibly aligned with the EU AI Act\footnote{Available at \url{https://www.europarl.europa.eu/news/en/headlines/society/20230601STO93804/eu-ai-act-first-regulation-on-artificial-intelligence}}. 
    
    Researchers can either build on past research or engage in investigations of the context by using interviews (e.g., with users, developers), observations (of the interaction) or secondary data of interactions. 
\\
     \midrule  
    
     \multirow{2}{3.5cm}{(4) Integrating context-specific factors into the theory}
     & 
    \textit{How can context-specific factors best be integrated into the theory?}\\\cmidrule{2-2} 

     &Identify the best way to include context-specific factors, by exploring how they relate to the pre-existing core constructs of the theory. For instance, considering the constructs of perceived ability, integrity, and benevolence from \citet{mayer-et_al_1995}, start by identifying the unique factors that are relevant to the specific context in which the theory is being applied. These could be cultural, organizational, situational, or demographic factors that are not originally accounted for. If the context-specific elements are distinct from the theoretical constructs (e.g., the situational context), and should be modelled as new elements within the theory, research should consider:\\
    &(a) including context-specific new factors that directly influence the core constructs (e.g.,  ability, integrity, and benevolence) in the theory. In this case, the core relationships specified in the theory would remain unchanged, but independent or dependent variables provide additional explanatory power;\\
    &(b) including context-specific factors as moderators of the relationship propagated in the theory. In this case, a relationship between variables would be altered by the contextual factor.
    \end{tabular}
    \caption{Guidelines for context-specific theorizing, adapted from \citep{Hong2014AFA} and tailored to conducting empirical research on trust in AI.}
    \label{tab:guidelines-theory}
\end{table}


\begin{thebibliography}{}
\providecommand{\doi}[1]{\url{https://doi.org/#1}}
\bibcommenthead

\bibitem[\protect\citeauthoryear{Aria and Cuccurullo}{Aria and
  Cuccurullo}{2017}]{Aria2017bibliometrixAR}
Aria, M. and C.~Cuccurullo. 2017.
\newblock Bibliometrix: {A}n {R}-tool for comprehensive science mapping
  analysis.
\newblock {\em J. Informetrics\/}~11: 959--975 .

\bibitem[\protect\citeauthoryear{Aria, Misuraca, and Spano}{Aria
  et~al.}{2020}]{Aria2020MappingTE}
Aria, M., M.~Misuraca, and M.~Spano. 2020.
\newblock Mapping the evolution of social research and data science on 30 years
  of social indicators research.
\newblock {\em Social Indicators Research\/}~149: 803 -- 831 .

\bibitem[\protect\citeauthoryear{Arici, Yildirim, Şeyma Caliklar, and
  Yilmaz}{Arici et~al.}{2019}]{Arici_2019}
Arici, F., P.~Yildirim, Şeyma Caliklar, and R.M. Yilmaz. 2019.
\newblock Research trends in the use of augmented reality in science education:
  Content and bibliometric mapping analysis.
\newblock {\em Computers \& Education\/}~142: 103647.
\newblock \doi{https://doi.org/10.1016/j.compedu.2019.103647} .

\bibitem[\protect\citeauthoryear{Babel, Kraus, Hock, Asenbauer, and
  Baumann}{Babel et~al.}{2021}]{Babel2021InvestigatingTV}
Babel, F., J.~Kraus, P.~Hock, H.~Asenbauer, and M.~Baumann 2021.
\newblock Investigating the validity of online robot evaluations: {C}omparison
  of findings from an one-sample online and laboratory study.
\newblock In {\em Companion of the 2021 ACM/IEEE International Conference on
  Human-Robot Interaction}, HRI '21 Companion, New York, NY, USA, pp.\
  116–120. Association for Computing Machinery.

\bibitem[\protect\citeauthoryear{Bansal, Bu{\c{c}}inca, Holstein, Hullman,
  Smith-Renner, Stumpf, and Wu}{Bansal et~al.}{2023}]{TRAIT}
Bansal, G., Z.~Bu{\c{c}}inca, K.~Holstein, J.~Hullman, A.M. Smith-Renner,
  S.~Stumpf, and S.~Wu 2023.
\newblock Workshop on trust and reliance in ai-human teams (trait).
\newblock In {\em Extended Abstracts of the 2023 CHI Conference on Human
  Factors in Computing Systems}, pp.\  1--6.

\bibitem[\protect\citeauthoryear{Bartneck, Kuli{\'c}, Croft, and
  Zoghbi}{Bartneck et~al.}{2009}]{Bartneck2009MeasurementIF}
Bartneck, C., D.~Kuli{\'c}, E.A. Croft, and S.~Zoghbi. 2009.
\newblock Measurement instruments for the anthropomorphism, animacy,
  likeability, perceived intelligence, and perceived safety of robots.
\newblock {\em International Journal of Social Robotics\/}~1: 71--81 .

\bibitem[\protect\citeauthoryear{Benk, Tolmeijer, von Wangenheim, and
  Ferrario}{Benk et~al.}{2022}]{benk2022value}
Benk, M., S.~Tolmeijer, F.~von Wangenheim, and A.~Ferrario. 2022.
\newblock The value of measuring trust in ai-{A} socio-technical system
  perspective.
\newblock {\em arXiv preprint arXiv:2204.13480\/} .

\bibitem[\protect\citeauthoryear{Bradford}{Bradford}{1985}]{Bradford1985SourcesOI}
Bradford, S.C. 1985.
\newblock Sources of information on specific subjects.
\newblock {\em Journal of Information Science\/}~10: 173--175 .

\bibitem[\protect\citeauthoryear{Brookes}{Brookes}{1969}]{Brookes1969BradfordsLA}
Brookes, B.C. 1969.
\newblock Bradford's law and the bibliography of science.
\newblock {\em Nature\/}~224: 953--956 .

\bibitem[\protect\citeauthoryear{Buccinca, Lin, Gajos, and Glassman}{Buccinca
  et~al.}{2020}]{Buccinca2020ProxyTA}
Buccinca, Z., P.~Lin, K.Z. Gajos, and E.L. Glassman. 2020.
\newblock Proxy tasks and subjective measures can be misleading in evaluating
  explainable {AI} systems.
\newblock {\em Proceedings of the 25th International Conference on Intelligent
  User Interfaces (IUI)\/}: 454--464 .

\bibitem[\protect\citeauthoryear{Bu{ç}inca, Malaya, and Gajos}{Bu{ç}inca
  et~al.}{2021}]{Buccinca_2021}
Bu{ç}inca, Z., M.B. Malaya, and K.Z. Gajos. 2021, apr.
\newblock To trust or to think: Cognitive forcing functions can reduce
  overreliance on ai in ai-assisted decision-making.
\newblock {\em Proc. ACM Hum.-Comput. Interact.\/}~{\em 5\/}(CSCW1).
\newblock \doi{10.1145/3449287} .

\bibitem[\protect\citeauthoryear{Cai, Reif, Hegde, Hipp, Kim, Smilkov,
  Wattenberg, Viegas, Corrado, Stumpe, and Terry}{Cai et~al.}{2019}]{Cai2019}
Cai, C.J., E.~Reif, N.~Hegde, J.~Hipp, B.~Kim, D.~Smilkov, M.~Wattenberg,
  F.~Viegas, G.S. Corrado, M.C. Stumpe, and M.~Terry. 2019.
\newblock Human-centered tools for coping with imperfect algorithms during
  medical decision-making.
\newblock pp.\  1–14.
\newblock \doi{10.1145/3290605.3300234} .

\bibitem[\protect\citeauthoryear{Castaldo, Premazzi, and Zerbini}{Castaldo
  et~al.}{2010}]{Castaldo2010TheMO}
Castaldo, S., K.~Premazzi, and F.~Zerbini. 2010.
\newblock The meaning(s) of trust. {A} content analysis on the diverse
  conceptualizations of trust in scholarly research on business relationships.
\newblock {\em Journal of Business Ethics\/}~96: 657--668 .

\bibitem[\protect\citeauthoryear{Castelfranchi and Falcone}{Castelfranchi and
  Falcone}{2010}]{Castelfranchi_Falcone_2010}
Castelfranchi, C. and R.~Falcone. 2010.
\newblock {\em Trust theory: {A} socio-cognitive and computational model\/}
  (1st ed.).
\newblock Wiley Publishing.

\bibitem[\protect\citeauthoryear{Cerovšek and Mikoš}{Cerovšek and
  Mikoš}{2014}]{CEROVSEK2014147}
Cerovšek, T. and M.~Mikoš. 2014.
\newblock A comparative study of cross-domain research output and citations:
  Research impact cubes and binary citation frequencies.
\newblock {\em Journal of Informetrics\/}~{\em 8\/}(1): 147--161.
\newblock \doi{https://doi.org/10.1016/j.joi.2013.11.004} .

\bibitem[\protect\citeauthoryear{Chi, Jia, Li, and G{\"u}rsoy}{Chi
  et~al.}{2021}]{Chi2021DevelopingAF}
Chi, O.H., S.~Jia, Y.~Li, and D.~G{\"u}rsoy. 2021.
\newblock Developing a formative scale to measure consumers' trust toward
  interaction with artificially intelligent (ai) social robots in service
  delivery.
\newblock {\em Comput. Hum. Behav.\/}~118: 106700 .

\bibitem[\protect\citeauthoryear{Chignell, Wang, Zare, and Li}{Chignell
  et~al.}{2022}]{Chignell2022TheEO}
Chignell, M.H., L.~Wang, A.~Zare, and J.J. Li. 2022.
\newblock The evolution of {HCI} and human factors: {I}ntegrating human and
  artificial intelligence.
\newblock {\em ACM Transactions on Computer-Human Interaction\/}~30: 1--30 .

\bibitem[\protect\citeauthoryear{Chiou and Lee}{Chiou and
  Lee}{2021}]{Chiou2021TrustingAD}
Chiou, E.K. and J.D. Lee. 2021.
\newblock Trusting automation: {D}esigning for responsivity and resilience.
\newblock {\em Human Factors: The Journal of Human Factors and Ergonomics
  Society\/}~65: 137--165 .

\bibitem[\protect\citeauthoryear{Choi, Lee, and Kim}{Choi
  et~al.}{2011}]{Choi2011TheIO}
Choi, J., H.J. Lee, and Y.C. Kim. 2011.
\newblock The influence of social presence on customer intention to reuse
  online recommender systems: {T}he roles of personalization and product type.
\newblock {\em International Journal of Electronic Commerce\/}~16: 129--154 .

\bibitem[\protect\citeauthoryear{Choi and Ji}{Choi and
  Ji}{2015}]{Choi2015InvestigatingTI}
Choi, J.K. and Y.G. Ji. 2015.
\newblock Investigating the importance of trust on adopting an autonomous
  vehicle.
\newblock {\em International Journal of Human-Computer Interaction\/}~31:
  692--702 .

\bibitem[\protect\citeauthoryear{Corallo, Latino, Menegoli, Devitiis, and
  Viscecchia}{Corallo et~al.}{2019}]{Corallo2019HumanFI}
Corallo, A., M.E. Latino, M.~Menegoli, B.D. Devitiis, and R.~Viscecchia. 2019.
\newblock Human factor in food label design to support consumer healthcare and
  safety: A systematic literature review.
\newblock {\em Sustainability\/} .

\bibitem[\protect\citeauthoryear{Culnan}{Culnan}{1986}]{Culnan1986TheID}
Culnan, M.J. 1986.
\newblock The intellectual development of management information systems,
  1972-1982: {A} co-citation analysis.
\newblock {\em Management Science\/}~32: 156--172 .

\bibitem[\protect\citeauthoryear{Davis}{Davis}{1989}]{Davis1989PerceivedUP}
Davis, F.D. 1989.
\newblock Perceived usefulness, perceived ease of use, and user acceptance of
  information technology.
\newblock {\em MIS Q.\/}~13: 319--340 .

\bibitem[\protect\citeauthoryear{Dekarske and Joshi}{Dekarske and
  Joshi}{2021}]{Dekarske2021HumanTO}
Dekarske, J. and S.S. Joshi. 2021.
\newblock Human trust of autonomous agent varies with strategy and capability
  in collaborative grid search task.
\newblock {\em 2021 IEEE 2nd International Conference on Human-Machine Systems
  (ICHMS)\/}: 1--6 .

\bibitem[\protect\citeauthoryear{Devlin, Chang, Lee, and Toutanova}{Devlin
  et~al.}{2018}]{devlin2018bert}
Devlin, J., M.W. Chang, K.~Lee, and K.~Toutanova. 2018.
\newblock Bert: {P}re-training of deep bidirectional transformers for language
  understanding.
\newblock {\em arXiv preprint arXiv:1810.04805\/} .

\bibitem[\protect\citeauthoryear{Dietvorst, Simmons, and Massey}{Dietvorst
  et~al.}{2018}]{dietvorst_overcoming_2018}
Dietvorst, B.J., J.P. Simmons, and C.~Massey. 2018.
\newblock Overcoming {Algorithm} {Aversion}: {People} {Will} {Use} {Imperfect}
  {Algorithms} {If} {They} {Can} ({Even} {Slightly}) {Modify} {Them}.
\newblock {\em Management Science\/}~{\em 64\/}(3): 1155--1170.
\newblock \doi{10.1287/mnsc.2016.2643} .

\bibitem[\protect\citeauthoryear{Djamasbi, Galletta, Nah, Page, Robert, and
  Wisniewski}{Djamasbi et~al.}{2018}]{Djamasbi2018BridgingAB}
Djamasbi, S., D.F. Galletta, F.F.H. Nah, X.~Page, L.P. Robert, and P.J.
  Wisniewski. 2018.
\newblock Bridging a bridge: Bringing two hci communities together.
\newblock {\em Extended Abstracts of the 2018 CHI Conference on Human Factors
  in Computing Systems\/} .

\bibitem[\protect\citeauthoryear{Dub{\'e} and Par{\'e}}{Dub{\'e} and
  Par{\'e}}{2003}]{Dub2003RigorII}
Dub{\'e}, L. and G.~Par{\'e}. 2003.
\newblock Rigor in information systems positivist case research: Current
  practices.
\newblock {\em MIS Q.\/}~27: 597--635 .

\bibitem[\protect\citeauthoryear{Dzindolet, Peterson, Pomranky, Pierce, and
  Beck}{Dzindolet et~al.}{2003}]{Dzindolet2003TheRO}
Dzindolet, M.T., S.A. Peterson, R.A. Pomranky, L.G. Pierce, and H.P. Beck.
  2003.
\newblock The role of trust in automation reliance.
\newblock {\em Int. J. Hum. Comput. Stud.\/}~58: 697--718 .

\bibitem[\protect\citeauthoryear{Ehsan and Riedl}{Ehsan and
  Riedl}{2020}]{Ehsan2020HumancenteredEA}
Ehsan, U. and M.O. Riedl 2020.
\newblock Human-centered explainable {AI}: Towards a reflective sociotechnical
  approach.
\newblock In {\em 2020 International Conference on Human-Computer Interaction}.

\bibitem[\protect\citeauthoryear{Enebechi and Duffy}{Enebechi and
  Duffy}{2020}]{Enebechi_2020}
Enebechi, C.N. and V.G. Duffy. 2020.
\newblock Virtual reality and artificial intelligence in mobile computing and
  applied ergonomics: A bibliometric and content analysis.
\newblock pp.\  334--345 .

\bibitem[\protect\citeauthoryear{Fahimnia, Sarkis, and Davarzani}{Fahimnia
  et~al.}{2015}]{FAHIMNIA2015101}
Fahimnia, B., J.~Sarkis, and H.~Davarzani. 2015.
\newblock Green supply chain management: A review and bibliometric analysis.
\newblock {\em International Journal of Production Economics\/}~162: 101--114.
\newblock \doi{https://doi.org/10.1016/j.ijpe.2015.01.003} .

\bibitem[\protect\citeauthoryear{Feher, Vicsek, and Deuze}{Feher
  et~al.}{2024}]{Feher2024ModelingAT}
Feher, K., L.~Vicsek, and M.~Deuze. 2024.
\newblock Modeling ai trust for 2050: perspectives from media and
  info-communication experts.
\newblock {\em AI \& SOCIETY\/}: 1--14 .

\bibitem[\protect\citeauthoryear{Ferrario and Loi}{Ferrario and
  Loi}{2022}]{ferrario2022explainability}
Ferrario, A. and M.~Loi 2022.
\newblock How explainability contributes to trust in ai.
\newblock In {\em Proceedings of the 2022 ACM Conference on Fairness,
  Accountability, and Transparency}, pp.\  1457--1466.

\bibitem[\protect\citeauthoryear{Ferrario, Loi, and Vigan\`o}{Ferrario
  et~al.}{2020}]{ferrario_ai_2019}
Ferrario, A., M.~Loi, and E.~Vigan\`o. 2020.
\newblock In {AI} we trust incrementally: {A} multi-layer model of trust to
  analyze human-artificial intelligence interactions.
\newblock {\em Philosophy \& Technology\/}~33: 523--539.
\newblock \doi{10.1007/s13347-019-00378-3} .

\bibitem[\protect\citeauthoryear{Freedy, de~Visser, Weltman, and
  Coeyman}{Freedy et~al.}{2007}]{Freedy2007MeasurementOT}
Freedy, A., E.~de~Visser, G.~Weltman, and N.~Coeyman. 2007.
\newblock Measurement of trust in human-robot collaboration.
\newblock {\em 2007 International Symposium on Collaborative Technologies and
  Systems\/}: 106--114 .

\bibitem[\protect\citeauthoryear{Gefen}{Gefen}{2013}]{gefen_psychology_2013}
Gefen, D. 2013.
\newblock {\em Psychology of trust: {N}ew research}.
\newblock Psychology of emotions, motivations and actions. New York: Nova
  Science Publishers.

\bibitem[\protect\citeauthoryear{Gillespie, Lockey, Curtis, Pool, and
  Akbari}{Gillespie et~al.}{2023}]{Gillespie2023TrustIA}
Gillespie, N.M., S.~Lockey, C.~Curtis, J.K. Pool, and A.~Akbari 2023.
\newblock Trust in artificial intelligence: A global study.

\bibitem[\protect\citeauthoryear{Glikson and Woolley}{Glikson and
  Woolley}{2020}]{Glikson2020HumanTI}
Glikson, E. and A.~Woolley. 2020.
\newblock Human trust in artificial intelligence: {R}eview of empirical
  research.
\newblock {\em The Academy of Management Annals\/}~14: 627--660 .

\bibitem[\protect\citeauthoryear{Guerreiro, Sato, Asakawa, Dong, Kitani, and
  Asakawa}{Guerreiro et~al.}{2019}]{Guerreiro2019CaBotDA}
Guerreiro, J.P.V., D.~Sato, S.~Asakawa, H.~Dong, K.M. Kitani, and C.~Asakawa.
  2019.
\newblock Cabot: {D}esigning and evaluating an autonomous navigation robot for
  blind people.
\newblock {\em Proceedings of the 21st International ACM SIGACCESS Conference
  on Computers and Accessibility\/}: 68--82 .

\bibitem[\protect\citeauthoryear{Guggemos, Seufert, and Sonderegger}{Guggemos
  et~al.}{2020}]{Guggemos2020HumanoidRI}
Guggemos, J., S.~Seufert, and S.~Sonderegger. 2020.
\newblock Humanoid robots in higher education: {E}valuating the acceptance of
  {P}epper in the context of an academic writing course using the {UTAUT}.
\newblock {\em Br. J. Educ. Technol.\/}~51: 1864--1883 .

\bibitem[\protect\citeauthoryear{Hancock, Billings, Schaefer, Chen, Visser, and
  Parasuraman}{Hancock et~al.}{2011}]{Hancock_2011}
Hancock, P.A., D.R. Billings, K.~Schaefer, J.~Chen, E.D. Visser, and
  R.~Parasuraman. 2011.
\newblock A meta-analysis of factors affecting trust in human-robot
  interaction.
\newblock {\em Human Factors: The Journal of Human Factors and Ergonomics
  Society\/}~53: 517--527 .

\bibitem[\protect\citeauthoryear{Hayashi and Wakabayashi}{Hayashi and
  Wakabayashi}{2017}]{Hayashi-et-al_2017}
Hayashi, Y. and K.~Wakabayashi 2017.
\newblock Can ai become reliable source to support human decision making in a
  court scene?
\newblock In {\em Companion of the 2017 ACM Conference on Computer Supported
  Cooperative Work and Social Computing}, CSCW '17 Companion, New York, NY,
  USA, pp.\  195–198. Association for Computing Machinery.

\bibitem[\protect\citeauthoryear{Hegner, Beldad, and Brunswick}{Hegner
  et~al.}{2019}]{Hegner2019InAW}
Hegner, S.M., A.~Beldad, and G.J. Brunswick. 2019.
\newblock In automatic we trust: {I}nvestigating the impact of trust, control,
  personality characteristics, and extrinsic and intrinsic motivations on the
  acceptance of autonomous vehicles.
\newblock {\em International Journal of Human–Computer Interaction\/}~35:
  1769--1780 .

\bibitem[\protect\citeauthoryear{Hirsch}{Hirsch}{2005}]{HirschIndex}
Hirsch, J.E. 2005.
\newblock An index to quantify an individual's scientific research output.
\newblock {\em Proceedings of the National Academy of Sciences\/}~{\em
  102\/}(46): 16569--16572.
\newblock \doi{10.1073/pnas.0507655102}.
\newblock
  {\href{https://arxiv.org/abs/https://www.pnas.org/doi/pdf/10.1073/pnas.0507655102}{{https://www.pnas.org/doi/pdf/10.1073/pnas.0507655102}}}
  .

\bibitem[\protect\citeauthoryear{Hoegen, Aneja, McDuff, and Czerwinski}{Hoegen
  et~al.}{2019}]{Hoegen2019AnEC}
Hoegen, R., D.~Aneja, D.J. McDuff, and M.~Czerwinski. 2019.
\newblock An end-to-end conversational style matching agent.
\newblock {\em Proceedings of the 19th ACM International Conference on
  Intelligent Virtual Agents\/}: 111--118 .

\bibitem[\protect\citeauthoryear{Hoff and Bashir}{Hoff and
  Bashir}{2015}]{Hoff2015TrustIA}
Hoff, K.A. and M.~Bashir. 2015.
\newblock Trust in automation: {I}ntegrating empirical evidence on factors that
  influence trust.
\newblock {\em Human Factors\/}~{\em 57\/}(3): 407--434.
\newblock \doi{10.1177/0018720814547570} .

\bibitem[\protect\citeauthoryear{Hoffman}{Hoffman}{2017}]{Hoffman2017ATO}
Hoffman, R.R. 2017.
\newblock A taxonomy of emergent trusting in the human–machine relationship.

\bibitem[\protect\citeauthoryear{Hong, Chan, Thong, Chasalow, and Dhillon}{Hong
  et~al.}{2014}]{Hong2014AFA}
Hong, W., F.K.Y. Chan, J.Y.L. Thong, L.C. Chasalow, and G.~Dhillon. 2014.
\newblock A framework and guidelines for context-specific theorizing in
  information systems research.
\newblock {\em Information Systems: Behavioral \& Social Methods eJournal\/} .

\bibitem[\protect\citeauthoryear{Inbar and Meyer}{Inbar and
  Meyer}{2019}]{Inbar2019PolitenessCP}
Inbar, O. and J.~Meyer. 2019.
\newblock Politeness counts: {P}erceptions of peacekeeping robots.
\newblock {\em IEEE Transactions on Human-Machine Systems\/}~49: 232--240 .

\bibitem[\protect\citeauthoryear{Jacobs, He, Pradier, Lam, Ahn, McCoy, Perlis,
  Doshi-Velez, and Gajos}{Jacobs et~al.}{2021}]{Jacobs2021DesigningAF}
Jacobs, M.L., J.~He, M.F. Pradier, B.~Lam, A.C. Ahn, T.H. McCoy, R.H. Perlis,
  F.~Doshi-Velez, and K.Z. Gajos. 2021.
\newblock Designing {AI} for trust and collaboration in time-constrained
  medical decisions: {A} sociotechnical lens.
\newblock {\em Proceedings of the 2021 CHI Conference on Human Factors in
  Computing Systems\/}: 1--14 .

\bibitem[\protect\citeauthoryear{Jacovi, Marasovi{\'c}, Miller, and
  Goldberg}{Jacovi et~al.}{2021}]{Jacovi2020FormalizingTI}
Jacovi, A., A.~Marasovi{\'c}, T.~Miller, and Y.~Goldberg. 2021.
\newblock Formalizing trust in artificial intelligence: {P}rerequisites, causes
  and goals of human trust in {AI}.
\newblock {\em Proceedings of the 2021 ACM Conference on Fairness,
  Accountability, and Transparency\/}: 624--635 .

\bibitem[\protect\citeauthoryear{Jian, Bisantz, Drury, and Llinas}{Jian
  et~al.}{2000}]{Jian2000FoundationsFA}
Jian, J.Y., A.~Bisantz, C.~Drury, and J.~Llinas. 2000.
\newblock Foundations for an empirically determined scale of trust in automated
  systems.
\newblock {\em International Journal of Cognitive Ergonomics\/}~4: 53--71 .

\bibitem[\protect\citeauthoryear{Jobin, Ienca, and Vayena}{Jobin
  et~al.}{2019}]{jobin2019global}
Jobin, A., M.~Ienca, and E.~Vayena. 2019.
\newblock The global landscape of {AI} ethics guidelines.
\newblock {\em Nature Machine Intelligence\/}~{\em 1\/}(9): 389--399 .

\bibitem[\protect\citeauthoryear{Kaur, Uslu, Rittichier, and Durresi}{Kaur
  et~al.}{2022}]{kaur2022trustworthy}
Kaur, D., S.~Uslu, K.J. Rittichier, and A.~Durresi. 2022.
\newblock Trustworthy artificial intelligence: {A} review.
\newblock {\em ACM Computing Surveys (CSUR)\/}~{\em 55\/}(2): 1--38 .

\bibitem[\protect\citeauthoryear{Kaur, Nori, Jenkins, Caruana, Wallach, and
  Wortman~Vaughan}{Kaur et~al.}{2020}]{kaur-et-al_2020}
Kaur, H., H.~Nori, S.~Jenkins, R.~Caruana, H.~Wallach, and J.~Wortman~Vaughan
  2020.
\newblock Interpreting interpretability: {U}nderstanding data scientists' use
  of interpretability tools for machine learning.
\newblock In {\em Proceedings of the 2020 CHI Conference on Human Factors in
  Computing Systems}, New York, NY, USA, pp.\  1--14.

\bibitem[\protect\citeauthoryear{Kenny, Ford, Quinn, and Keane}{Kenny
  et~al.}{2021}]{Kenny2021ExplainingBC}
Kenny, E.M., C.~Ford, M.S. Quinn, and M.~Keane. 2021.
\newblock Explaining black-box classifiers using post-hoc
  explanations-by-example: {T}he effect of explanations and error-rates in
  {XAI} user studies.
\newblock {\em Artif. Intell.\/}~294: 103459 .

\bibitem[\protect\citeauthoryear{Kim, Park, and Sundar}{Kim
  et~al.}{2013}]{Kim2013CaregivingRI}
Kim, K.J., E.~Park, and S.S. Sundar. 2013.
\newblock Caregiving role in human-robot interaction: {A} study of the
  mediating effects of perceived benefit and social presence.
\newblock {\em Comput. Hum. Behav.\/}~29: 1799--1806 .

\bibitem[\protect\citeauthoryear{Knickrehm, Voss, and Barton}{Knickrehm
  et~al.}{2023}]{knickrehm2023can}
Knickrehm, C., M.~Voss, and M.C. Barton. 2023.
\newblock Can you trust me? {U}sing {AI} to review more than three decades of
  ai trust literature.
\newblock {\em ECIS 2023 Research Papers\/}~302 .

\bibitem[\protect\citeauthoryear{Knowles and Richards}{Knowles and
  Richards}{2021}]{Knowles2021TheSO}
Knowles, B. and J.T. Richards. 2021.
\newblock The sanction of authority: Promoting public trust in ai.
\newblock {\em Proceedings of the 2021 ACM Conference on Fairness,
  Accountability, and Transparency\/} .

\bibitem[\protect\citeauthoryear{Kreps, George, Lushenko, and Rao}{Kreps
  et~al.}{2023}]{Kreps2023ExploringTA}
Kreps, S., J.~George, P.~Lushenko, and A.B. Rao. 2023.
\newblock Exploring the artificial intelligence “trust paradox”: Evidence
  from a survey experiment in the united states.
\newblock {\em PLOS ONE\/}~18 .

\bibitem[\protect\citeauthoryear{Kroeger}{Kroeger}{2016}]{Kroeger2016FaceworkCT}
Kroeger, F. 2016.
\newblock Facework: creating trust in systems, institutions and organisations.
\newblock {\em Cambridge Journal of Economics\/}~41: 487--514 .

\bibitem[\protect\citeauthoryear{Langer, Hunsicker, Feldkamp, K{\"o}nig, and
  Grgi{\'c}-Hlača}{Langer et~al.}{2022}]{Langer2022}
Langer, M., T.~Hunsicker, T.~Feldkamp, C.J. K{\"o}nig, and N.~Grgi{\'c}-Hlača.
  2022.
\newblock ``look! it’s a computer program! {I}t’s an algorithm! {I}t’s
  {AI}!'': {D}oes terminology affect human perceptions and evaluations of
  algorithmic decision-making systems?
\newblock {\em Proceedings of the 2022 CHI Conference on Human Factors in
  Computing Systems\/}: 1--28.
\newblock \doi{https://doi.org/10.1145/3491102.3517527} .

\bibitem[\protect\citeauthoryear{Laux, Wachter, and Mittelstadt}{Laux
  et~al.}{2023}]{Laux2023TrustworthyAI}
Laux, J., S.~Wachter, and B.D. Mittelstadt. 2023.
\newblock Trustworthy artificial intelligence and the european union ai act: On
  the conflation of trustworthiness and acceptability of risk.
\newblock {\em Regulation \& Governance\/}~18: 3 -- 32 .

\bibitem[\protect\citeauthoryear{Laxar, Eitenberger, Maleczek, Kaider,
  Hammerle, and Kimberger}{Laxar et~al.}{2023}]{Laxar-et-al_2023}
Laxar, D., M.~Eitenberger, M.~Maleczek, A.~Kaider, F.~Hammerle, and
  O.~Kimberger. 2023, 09.
\newblock The influence of explainable vs non-explainable clinical decision
  support systems on rapid triage decisions: a mixed methods study.
\newblock {\em BMC Medicine\/}~21.
\newblock \doi{10.1186/s12916-023-03068-2} .

\bibitem[\protect\citeauthoryear{Lee and See}{Lee and
  See}{2004}]{lee_trust_2004}
Lee, J.D. and K.A. See. 2004.
\newblock Trust in automation: {D}esigning for appropriate reliance.
\newblock {\em Human factors\/}~{\em 46\/}(1): 50--80 .

\bibitem[\protect\citeauthoryear{Leichtmann, Humer, Hinterreiter, Streit, and
  Mara}{Leichtmann et~al.}{2022}]{Leichtmann2022EffectsOE}
Leichtmann, B., C.~Humer, A.P. Hinterreiter, M.~Streit, and M.~Mara. 2022.
\newblock Effects of explainable artificial intelligence on trust and human
  behavior in a high-risk decision task.
\newblock {\em Comput. Hum. Behav.\/}~139: 107539 .

\bibitem[\protect\citeauthoryear{Li, Qi, Liu, Di, Liu, Pei, Yi, and Zhou}{Li
  et~al.}{2023}]{li2021trustworthy}
Li, B., P.~Qi, B.~Liu, S.~Di, J.~Liu, J.~Pei, J.~Yi, and B.~Zhou. 2023.
\newblock Trustworthy {AI}: {F}rom principles to practices.
\newblock {\em ACM Computing Surveys\/}~{\em 55\/}(9): 1--46 .

\bibitem[\protect\citeauthoryear{Linxen, Sturm, Br{\"u}hlmann, Cassau, Opwis,
  and Reinecke}{Linxen et~al.}{2021}]{Linxen2021HowWI}
Linxen, S., C.~Sturm, F.~Br{\"u}hlmann, V.~Cassau, K.~Opwis, and K.~Reinecke.
  2021.
\newblock How weird is chi?
\newblock {\em Proceedings of the 2021 CHI Conference on Human Factors in
  Computing Systems\/} .

\bibitem[\protect\citeauthoryear{Liu, Gonçalves, Ferreira, Xiao, Hosio, and
  Kostakos}{Liu et~al.}{2014}]{Liu2014CHI1M}
Liu, Y., J.~Gonçalves, D.~Ferreira, B.~Xiao, S.J. Hosio, and V.~Kostakos.
  2014.
\newblock Chi 1994-2013: {M}apping two decades of intellectual progress through
  co-word analysis.
\newblock {\em Proceedings of the SIGCHI Conference on Human Factors in
  Computing Systems\/}: 3553--3562 .

\bibitem[\protect\citeauthoryear{Loi, Ferrario, and Vigan{\`o}}{Loi
  et~al.}{2023}]{loi2023much}
Loi, M., A.~Ferrario, and E.~Vigan{\`o}. 2023.
\newblock How much do you trust me? a logico-mathematical analysis of the
  concept of the intensity of trust.
\newblock {\em Synthese\/}~{\em 201\/}(6): 186 .

\bibitem[\protect\citeauthoryear{Lorenz, Nocera, and Parasuraman}{Lorenz
  et~al.}{2002}]{Lorenz2002DisplayIE}
Lorenz, B., F.D. Nocera, and R.~Parasuraman. 2002.
\newblock Display integration enhances information sampling and decision making
  in automated fault management in a simulated spaceflight micro-world.
\newblock {\em Proceedings of the Human Factors and Ergonomics Society Annual
  Meeting\/}~46: 31 -- 35 .

\bibitem[\protect\citeauthoryear{Lucaj, van~der Smagt, and Benbouzid}{Lucaj
  et~al.}{2023}]{Lucaj2023AIRI}
Lucaj, L., P.~van~der Smagt, and D.~Benbouzid. 2023.
\newblock Ai regulation is (not) all you need.
\newblock {\em Proceedings of the 2023 ACM Conference on Fairness,
  Accountability, and Transparency\/}: 1267--1279 .

\bibitem[\protect\citeauthoryear{Mayer, Davis, and Schoorman}{Mayer
  et~al.}{1995}]{mayer-et_al_1995}
Mayer, R.C., J.H. Davis, and F.D. Schoorman. 1995.
\newblock An integrative model of organizational trust.
\newblock {\em Academy of Management Review\/}~20: 709--734 .

\bibitem[\protect\citeauthoryear{McKnight and Chervany}{McKnight and
  Chervany}{1996}]{mcknight_chervany_1996}
McKnight, D. and N.~Chervany. 1996.
\newblock {\em The Meanings of Trust}.
\newblock Minneapolis, Minn., USA: Carlson School of Management, Univ. of
  Minnesota.

\bibitem[\protect\citeauthoryear{Mcknight, Carter, Thatcher, and Clay}{Mcknight
  et~al.}{2011}]{mcknight_trust_2011}
Mcknight, D.H., M.~Carter, J.B. Thatcher, and P.F. Clay. 2011, June.
\newblock Trust in a specific technology: {An} investigation of its components
  and measures.
\newblock {\em ACM Transactions on Management Information Systems\/}~{\em
  2\/}(2): 1--25.
\newblock \doi{10.1145/1985347.1985353} .

\bibitem[\protect\citeauthoryear{McKnight and Chervany}{McKnight and
  Chervany}{2000}]{McKnight2000WhatIT}
McKnight, D.H. and N.L. Chervany 2000.
\newblock What is trust? {A} conceptual analysis and an interdisciplinary
  model.
\newblock pp.\  282.

\bibitem[\protect\citeauthoryear{Meidute-Kavaliauskiene, Y{\i}ld{\i}z,
  {\c{C}}i{\u{g}}dem, and {\v{C}}in{\v{c}}ikait{\.e}}{Meidute-Kavaliauskiene
  et~al.}{2021}]{MeidutKavaliauskien2021TheEO}
Meidute-Kavaliauskiene, I., B.~Y{\i}ld{\i}z, {\c{S}}.~{\c{C}}i{\u{g}}dem, and
  R.~{\v{C}}in{\v{c}}ikait{\.e}. 2021.
\newblock The effect of covid-19 on airline transportation services: {A} study
  on service robot usage intention.
\newblock {\em Sustainability\/}~{\em 13\/}(22): 12571 .

\bibitem[\protect\citeauthoryear{Mercado, Rupp, Chen, Barnes, Barber, and
  Procci}{Mercado et~al.}{2016}]{Mercado2016IntelligentAT}
Mercado, J.E., M.A. Rupp, J.Y. Chen, M.J. Barnes, D.J. Barber, and K.~Procci.
  2016.
\newblock Intelligent agent transparency in human–agent teaming for
  multi-{UxV} management.
\newblock {\em Human Factors: The Journal of Human Factors and Ergonomics
  Society\/}~58: 401--415 .

\bibitem[\protect\citeauthoryear{Miller}{Miller}{2019}]{miller_explanation_2018}
Miller, T. 2019.
\newblock Explanation in artificial intelligence: Insights from the social
  sciences.
\newblock {\em Artificial Intelligence\/}~267: 1--38 .

\bibitem[\protect\citeauthoryear{Nadarzynski, Miles, Cowie, and
  Ridge}{Nadarzynski et~al.}{2019}]{Nadarzynski2019AcceptabilityOA}
Nadarzynski, T., O.~Miles, A.~Cowie, and D.T. Ridge. 2019.
\newblock Acceptability of artificial intelligence (ai)-led chatbot services in
  healthcare: A mixed-methods study.
\newblock {\em Digital Health\/}~5 .

\bibitem[\protect\citeauthoryear{Nass, Steuer, and Tauber}{Nass
  et~al.}{1994}]{Nass1994ComputersAS}
Nass, C., J.~Steuer, and E.R. Tauber 1994.
\newblock Computers are social actors.
\newblock In {\em International Conference on Human Factors in Computing
  Systems}, pp.\  72--78.

\bibitem[\protect\citeauthoryear{Natarajan and Gombolay}{Natarajan and
  Gombolay}{2020}]{Natarajan2020EffectsOA}
Natarajan, M. and M.C. Gombolay. 2020.
\newblock Effects of anthropomorphism and accountability on trust in human
  robot interaction.
\newblock {\em 2020 15th ACM/IEEE International Conference on Human-Robot
  Interaction (HRI)\/}: 33--42 .

\bibitem[\protect\citeauthoryear{Nerur, Rasheed, and Natarajan}{Nerur
  et~al.}{2008}]{Nerur2008TheIS}
Nerur, S.P., A.A. Rasheed, and V.~Natarajan. 2008.
\newblock The intellectual structure of the strategic management field: {A}n
  author co‐citation analysis.
\newblock {\em Southern Medical Journal\/}~29: 319--336 .

\bibitem[\protect\citeauthoryear{Nielsen and Andersen}{Nielsen and
  Andersen}{2021}]{Nielsen2021}
Nielsen, M.W. and J.P. Andersen. 2021.
\newblock Global citation inequality is on the rise.
\newblock {\em Proceedings of the National Academy of Sciences\/}~{\em
  118\/}(7): e2012208118.
\newblock \doi{10.1073/pnas.2012208118}.
\newblock
  {\href{https://arxiv.org/abs/https://www.pnas.org/doi/pdf/10.1073/pnas.2012208118}{{https://www.pnas.org/doi/pdf/10.1073/pnas.2012208118}}}
  .

\bibitem[\protect\citeauthoryear{Nikolaidis, Lasota, Ramakrishnan, and
  Shah}{Nikolaidis et~al.}{2015}]{Nikolaidis2015ImprovedHT}
Nikolaidis, S., P.A. Lasota, R.~Ramakrishnan, and J.A. Shah. 2015.
\newblock Improved human–robot team performance through cross-training, an
  approach inspired by human team training practices.
\newblock {\em The International Journal of Robotics Research\/}~34: 1711--1730
  .

\bibitem[\protect\citeauthoryear{Page, McKenzie, Bossuyt, Boutron, Hoffmann,
  Mulrow, Shamseer, Tetzlaff, Akl, Brennan, et~al.}{Page
  et~al.}{2021}]{Page2020TheP2}
Page, M.J., J.E. McKenzie, P.M. Bossuyt, I.~Boutron, T.C. Hoffmann, C.D.
  Mulrow, L.~Shamseer, J.M. Tetzlaff, E.A. Akl, S.E. Brennan, et~al. 2021.
\newblock The prisma 2020 statement: {A}n updated guideline for reporting
  systematic reviews.
\newblock {\em International journal of surgery\/}~88: 105906 .

\bibitem[\protect\citeauthoryear{Panagiotopoulos and
  Dimitrakopoulos}{Panagiotopoulos and
  Dimitrakopoulos}{2018}]{Panagiotopoulos2018AnEI}
Panagiotopoulos, I.E. and G.J. Dimitrakopoulos. 2018.
\newblock An empirical investigation on consumers’ intentions towards
  autonomous driving.
\newblock {\em Transportation Research Part C: Emerging Technologies\/}~95:
  773--784 .

\bibitem[\protect\citeauthoryear{Papagni, de~Pagter, Zafari, Filzmoser, and
  Koeszegi}{Papagni et~al.}{2022}]{Papagni2022ArtificialAE}
Papagni, G., J.~de~Pagter, S.~Zafari, M.~Filzmoser, and S.T. Koeszegi. 2022.
\newblock Artificial agents’ explainability to support trust: considerations
  on timing and context.
\newblock {\em AI \& SOCIETY\/}~38: 947--960 .

\bibitem[\protect\citeauthoryear{Parasuraman and Riley}{Parasuraman and
  Riley}{1997}]{Parasuraman1997HumansAA}
Parasuraman, R. and V.A. Riley. 1997.
\newblock Humans and automation: {U}se, misuse, disuse, abuse.
\newblock {\em Human Factors: The Journal of Human Factors and Ergonomics
  Society\/}~39: 230--253 .

\bibitem[\protect\citeauthoryear{Pavlou}{Pavlou}{2003}]{pavlou2003consumer}
Pavlou, P.A. 2003.
\newblock Consumer acceptance of electronic commerce: Integrating trust and
  risk with the {T}echnology {A}cceptance {M}odel.
\newblock {\em International journal of electronic commerce\/}~{\em 7\/}(3):
  101--134 .

\bibitem[\protect\citeauthoryear{Petersen, Potdevin, Mohammadi, Zidowitz,
  Breyer, Nowotka, Henn, Pechmann, Leucker, Rostalski, et~al.}{Petersen
  et~al.}{2022}]{petersen2022responsible}
Petersen, E., Y.~Potdevin, E.~Mohammadi, S.~Zidowitz, S.~Breyer, D.~Nowotka,
  S.~Henn, L.~Pechmann, M.~Leucker, P.~Rostalski, et~al. 2022.
\newblock Responsible and regulatory conform machine learning for medicine: A
  survey of challenges and solutions.
\newblock {\em IEEE Access\/}~10: 58375--58418 .

\bibitem[\protect\citeauthoryear{Pillai and Sivathanu}{Pillai and
  Sivathanu}{2020}]{Pillai2020AdoptionOA}
Pillai, R. and B.~Sivathanu. 2020.
\newblock Adoption of ai-based chatbots for hospitality and tourism.
\newblock {\em International Journal of Contemporary Hospitality Management\/}
  .

\bibitem[\protect\citeauthoryear{P{\"o}hler, Heine, and Deml}{P{\"o}hler
  et~al.}{2016}]{Phler2016ItemanalyseUF}
P{\"o}hler, G., T.~Heine, and B.~Deml. 2016.
\newblock Itemanalyse und faktorstruktur eines fragebogens zur messung von
  vertrauen im umgang mit automatischen systemen.
\newblock {\em Zeitschrift f{\"u}r Arbeitswissenschaft\/}~70: 151--160 .

\bibitem[\protect\citeauthoryear{Pu and Chen}{Pu and Chen}{2006}]{pu2006trust}
Pu, P. and L.~Chen 2006.
\newblock Trust building with explanation interfaces.
\newblock In {\em Proceedings of the 11th international conference on
  Intelligent user interfaces}, pp.\  93--100.

\bibitem[\protect\citeauthoryear{Pu and Chen}{Pu and
  Chen}{2007}]{Pu2007TrustinspiringEI}
Pu, P. and L.~Chen. 2007.
\newblock Trust-inspiring explanation interfaces for recommender systems.
\newblock {\em Knowl. Based Syst.\/}~20: 542--556 .

\bibitem[\protect\citeauthoryear{Rau, Li, and Li}{Rau
  et~al.}{2009a}]{Rau2009EffectsOC}
Rau, P.L.P., Y.~Li, and D.~Li. 2009a.
\newblock Effects of communication style and culture on ability to accept
  recommendations from robots.
\newblock {\em Comput. Hum. Behav.\/}~25: 587--595 .

\bibitem[\protect\citeauthoryear{Rau, Li, and Li}{Rau et~al.}{2009b}]{Rau_2009}
Rau, P.P., Y.~Li, and D.~Li. 2009b, mar.
\newblock Effects of communication style and culture on ability to accept
  recommendations from robots.
\newblock {\em Comput. Hum. Behav.\/}~{\em 25\/}(2): 587–595.
\newblock \doi{10.1016/j.chb.2008.12.025} .

\bibitem[\protect\citeauthoryear{Ribeiro, Singh, and Guestrin}{Ribeiro
  et~al.}{2016}]{Ribeiro-et-al_2016}
Ribeiro, M.T., S.~Singh, and C.~Guestrin 2016.
\newblock `{Why} should {I} trust you?': {E}xplaining the predictions of any
  classifier.
\newblock In {\em Proceedings of the 22nd ACM SIGKDD International Conference
  on Knowledge Discovery and Data Mining (KDD}, New York, NY, USA, pp.\
  1135--1144. Association for Computing Machinery.

\bibitem[\protect\citeauthoryear{Ridwan, Govindaraju, and Andriani}{Ridwan
  et~al.}{2023}]{Ridwan_22}
Ridwan, A.Y., R.~Govindaraju, and M.~Andriani. 2023.
\newblock Business analytic and business value: A review and bibliometric
  analysis of a decade of research.
\newblock pp.\  158–164.
\newblock \doi{10.1145/3568231.3568245} .

\bibitem[\protect\citeauthoryear{Rix}{Rix}{2022}]{Rix2022FromTT}
Rix, J. 2022.
\newblock From tools to teammates: Conceptualizing humans' perception of
  machines as teammates with a systematic literature review.
\newblock In {\em Hawaii International Conference on System Sciences}.

\bibitem[\protect\citeauthoryear{Robinette, Wagner, and Howard}{Robinette
  et~al.}{2016}]{Robinette2016}
Robinette, P., A.R. Wagner, and A.M. Howard 2016.
\newblock {\em Investigating human-robot trust in emergency scenarios:
  {M}ethodological lessons learned}, pp.\  143--166.
\newblock Boston, MA: Springer US.

\bibitem[\protect\citeauthoryear{Ruijten, Terken, and Chandramouli}{Ruijten
  et~al.}{2018}]{Ruijten2018EnhancingTI}
Ruijten, P.A.M., J.M.B. Terken, and S.N. Chandramouli. 2018.
\newblock Enhancing trust in autonomous vehicles through intelligent user
  interfaces that mimic human behavior.
\newblock {\em Multimodal Technol. Interact.\/}~2: 62 .

\bibitem[\protect\citeauthoryear{Salem, Lakatos, Amirabdollahian, and
  Dautenhahn}{Salem et~al.}{2015}]{Salem2015WouldYT}
Salem, M., G.~Lakatos, F.~Amirabdollahian, and K.~Dautenhahn 2015.
\newblock Would you trust a (faulty) robot? {E}ffects of error, task type and
  personality on human-robot cooperation and trust.
\newblock pp.\  141--148.

\bibitem[\protect\citeauthoryear{Salimzadeh, He, and Gadiraju}{Salimzadeh
  et~al.}{2023}]{Salimzadeh_2023}
Salimzadeh, S., G.~He, and U.~Gadiraju 2023.
\newblock A missing piece in the puzzle: Considering the role of task
  complexity in human-ai decision making.
\newblock In {\em Proceedings of the 31st ACM Conference on User Modeling,
  Adaptation and Personalization}, UMAP '23, New York, NY, USA, pp.\
  215–227. Association for Computing Machinery.

\bibitem[\protect\citeauthoryear{Salini}{Salini}{2016}]{introbibliometrix}
Salini, S. 2016.
\newblock {\em An introduction to bibliometrics}, Chapter~14, pp.\  130--143.
\newblock John Wiley \& Sons, Ltd.

\bibitem[\protect\citeauthoryear{Schemmer, Hemmer, K{\"u}hl, Benz, and
  Satzger}{Schemmer et~al.}{2022}]{schemmer2022should}
Schemmer, M., P.~Hemmer, N.~K{\"u}hl, C.~Benz, and G.~Satzger. 2022.
\newblock Should i follow ai-based advice? measuring appropriate reliance in
  human-ai decision-making.
\newblock {\em arXiv preprint arXiv:2204.06916\/} .

\bibitem[\protect\citeauthoryear{Schraagen, Elsasser, Fricke, Hof, and
  Ragalmuto}{Schraagen et~al.}{2020}]{Schraagen2020TrustingTX}
Schraagen, J.M.C., P.~Elsasser, H.~Fricke, M.~Hof, and F.~Ragalmuto. 2020.
\newblock Trusting the x in xai: Effects of different types of explanations by
  a self-driving car on trust, explanation satisfaction and mental models.
\newblock {\em Proceedings of the Human Factors and Ergonomics Society Annual
  Meeting\/}~64: 339--343 .

\bibitem[\protect\citeauthoryear{Shiffrin, B{\"o}rner, and Stigler}{Shiffrin
  et~al.}{2017}]{Shiffrin2017ScientificPD}
Shiffrin, R.M., K.~B{\"o}rner, and S.M. Stigler. 2017.
\newblock Scientific progress despite irreproducibility: A seeming paradox.
\newblock {\em Proceedings of the National Academy of Sciences\/}~115:
  2632--2639 .

\bibitem[\protect\citeauthoryear{Shin}{Shin}{2021}]{Shin2021TheEO}
Shin, D. 2021.
\newblock The effects of explainability and causability on perception, trust,
  and acceptance: Implications for explainable ai.
\newblock {\em Int. J. Hum. Comput. Stud.\/}~146: 102551 .

\bibitem[\protect\citeauthoryear{Simmler and Frischknecht}{Simmler and
  Frischknecht}{2021}]{simmler2021taxonomy}
Simmler, M. and R.~Frischknecht. 2021.
\newblock A taxonomy of human--machine collaboration: {C}apturing automation
  and technical autonomy.
\newblock {\em AI \& Society\/}~{\em 36\/}(1): 239--250 .

\bibitem[\protect\citeauthoryear{Slade, Dwivedi, Piercy, and Williams}{Slade
  et~al.}{2015}]{UTAUT}
Slade, E.L., Y.K. Dwivedi, N.C. Piercy, and M.D. Williams. 2015.
\newblock Modeling consumers’ adoption intentions of remote mobile payments
  in the united kingdom: {E}xtending {UTAUT} with innovativeness, risk and
  trust.
\newblock {\em Psychology \& Marketing\/}~{\em 32\/}(8): 860--873.
\newblock \doi{https://doi.org/10.1002/mar.20823}.
\newblock
  {\href{https://arxiv.org/abs/https://onlinelibrary.wiley.com/doi/pdf/10.1002/mar.20823}{{https://onlinelibrary.wiley.com/doi/pdf/10.1002/mar.20823}}}
  .

\bibitem[\protect\citeauthoryear{Spain, Bustamante, and Bliss}{Spain
  et~al.}{2008}]{Spain2008TowardsAE}
Spain, R.D., E.A. Bustamante, and J.P. Bliss. 2008.
\newblock Towards an empirically developed scale for system trust: Take two.
\newblock {\em Proceedings of the Human Factors and Ergonomics Society Annual
  Meeting\/}~52: 1335 -- 1339 .

\bibitem[\protect\citeauthoryear{Toreini, Aitken, Coopamootoo, Elliott, Zelaya,
  and van Moorsel}{Toreini et~al.}{2020}]{toreini_relationship_2019}
Toreini, E., M.~Aitken, K.~Coopamootoo, K.~Elliott, C.G. Zelaya, and A.~van
  Moorsel 2020.
\newblock The relationship between trust in {AI} and trustworthy machine
  learning technologies.
\newblock In {\em Proceedings of the 2020 Conference on Fairness,
  Accountability, and Transparency}, New York, NY, USA, pp.\  272--283.

\bibitem[\protect\citeauthoryear{van Eck and Waltman}{van Eck and
  Waltman}{2009}]{Eck2009SoftwareSV}
van Eck, N.J. and L.~Waltman. 2009.
\newblock Software survey: Vosviewer, a computer program for bibliometric
  mapping.
\newblock {\em Scientometrics\/}~84: 523 -- 538 .

\bibitem[\protect\citeauthoryear{Venkatesh, Thong, and Xu}{Venkatesh
  et~al.}{2012}]{Venkatesh2012ConsumerAA}
Venkatesh, V., J.Y.L. Thong, and X.~Xu. 2012.
\newblock Consumer acceptance and use of information technology: Extending the
  unified theory of acceptance and use of technology.
\newblock {\em Behavioral Marketing eJournal\/} .

\bibitem[\protect\citeauthoryear{Vereschak, Bailly, and Caramiaux}{Vereschak
  et~al.}{2021}]{Vereschak2021HowTE}
Vereschak, O., G.~Bailly, and B.~Caramiaux. 2021.
\newblock How to evaluate trust in {AI}-assisted decision making? {A} survey of
  empirical methodologies.
\newblock {\em Proceedings of the ACM on Human-Computer Interaction\/}~5: 1--39
  .

\bibitem[\protect\citeauthoryear{Wamba, Bawack, Guthrie, Queiroz, and
  Carillo}{Wamba et~al.}{2021}]{Wamba2020AreWP}
Wamba, S.F., R.E. Bawack, C.~Guthrie, M.M. Queiroz, and K.D.A. Carillo. 2021.
\newblock Are we preparing for a good {AI} society? {A} bibliometric review and
  research agenda.
\newblock {\em Technological Forecasting and Social Change\/}~164: 120482 .

\bibitem[\protect\citeauthoryear{Wang, Yu, Hu, and Li}{Wang
  et~al.}{2020}]{Wang2020ParticipantOS}
Wang, S., H.~Yu, X.~Hu, and J.~Li. 2020.
\newblock Participant or spectator? {C}omprehending the willingness of faculty
  to use intelligent tutoring systems in the artificial intelligence era.
\newblock {\em Br. J. Educ. Technol.\/}~51: 1657--1673 .

\bibitem[\protect\citeauthoryear{Waytz, Heafner, and Epley}{Waytz
  et~al.}{2014}]{Waytz2014TheMI}
Waytz, A., J.~Heafner, and N.~Epley. 2014.
\newblock The mind in the machine: {A}nthropomorphism increases trust in an
  autonomous vehicle.
\newblock {\em Journal of Experimental Social Psychology\/}~52: 113--117 .

\bibitem[\protect\citeauthoryear{Wu, Liu, and Chu}{Wu
  et~al.}{2023}]{Wu2023HowII}
Wu, W., R.~Liu, and J.~Chu. 2023.
\newblock How important is trust: Exploring the factors influencing college
  students' use of chat gpt as a learning aid.
\newblock {\em 2023 16th International Symposium on Computational Intelligence
  and Design (ISCID)\/}: 67--70 .

\bibitem[\protect\citeauthoryear{Yang, Huang, Scholtz, and Arendt}{Yang
  et~al.}{2020}]{Yang2020HowDV}
Yang, F., Z.~Huang, J.~Scholtz, and D.L. Arendt 2020.
\newblock How do visual explanations foster end users' appropriate trust in
  machine learning?
\newblock In {\em Proceedings of the 25th International Conference on
  Intelligent User Interfaces}, IUI '20, New York, NY, USA, pp.\  189–201.
  Association for Computing Machinery.

\bibitem[\protect\citeauthoryear{Yang, Zeng, Peng, and Jiang}{Yang
  et~al.}{2019}]{Yang2019AttitudesOC}
Yang, K., Z.~Zeng, H.~Peng, and Y.~Jiang. 2019.
\newblock Attitudes of chinese cancer patients toward the clinical use of
  artificial intelligence.
\newblock {\em Patient preference and adherence\/}~13: 1867--1875 .

\bibitem[\protect\citeauthoryear{Yang, Unhelkar, Li, and Shah}{Yang
  et~al.}{2017}]{Yang2017}
Yang, X.J., V.V. Unhelkar, K.~Li, and J.A. Shah. 2017.
\newblock Evaluating effects of user experience and system transparency on
  trust in automation.
\newblock pp.\  408–416.
\newblock \doi{10.1145/2909824.3020230} .

\bibitem[\protect\citeauthoryear{Yin, Wortman~Vaughan, and Wallach}{Yin
  et~al.}{2019}]{Yin2019UnderstandingTE}
Yin, M., J.~Wortman~Vaughan, and H.~Wallach 2019.
\newblock Understanding the effect of accuracy on trust in machine learning
  models.
\newblock In {\em Proceedings of the 2019 CHI Conference on Human Factors in
  Computing Systems}, CHI '19, New York, NY, USA, pp.\  1–12. Association for
  Computing Machinery.

\bibitem[\protect\citeauthoryear{Yu, Rosenfeld, and Gupta}{Yu
  et~al.}{2023}]{World-Economic-Forum}
Yu, D., H.~Rosenfeld, and A.~Gupta. 2023.
\newblock The ‘ai divide’ between the {G}lobal {N}orth and the {G}lobal
  {S}outh.
\newblock Accessed: 2022-07-20.

\bibitem[\protect\citeauthoryear{Zanker}{Zanker}{2012}]{Zanker2012TheIO}
Zanker, M. 2012.
\newblock The influence of knowledgeable explanations on users' perception of a
  recommender system.
\newblock {\em Proceedings of the sixth ACM conference on Recommender
  systems\/}: 269--272 .

\bibitem[\protect\citeauthoryear{Zhang, Pentina, and Fan}{Zhang
  et~al.}{2021}]{Zhang-et-al_2021_Marketing}
Zhang, L., I.~Pentina, and Y.~Fan. 2021, 01.
\newblock Who do you choose? comparing perceptions of human vs. robo-advisor in
  the context of financial services.
\newblock {\em Journal of Services Marketing\/}~ahead-of-print.
\newblock \doi{10.1108/JSM-05-2020-0162} .

\bibitem[\protect\citeauthoryear{Zhang, Liao, and Bellamy}{Zhang
  et~al.}{2020}]{Zhang-et-al_2020}
Zhang, Y., Q.V. Liao, and R.K.E. Bellamy 2020.
\newblock Effect of confidence and explanation on accuracy and trust
  calibration in {AI}-assisted decision making.
\newblock In {\em Proceedings of the 2020 Conference on Fairness,
  Accountability, and Transparency}, New York, NY, USA, pp.\  295--305.

\bibitem[\protect\citeauthoryear{Zhao, Ren, and Cheah}{Zhao
  et~al.}{2023}]{Zhao2023LeadingVR}
Zhao, X., Y.~Ren, and K.S.L. Cheah. 2023.
\newblock Leading virtual reality (vr) and augmented reality (ar) in education:
  Bibliometric and content analysis from the web of science (2018–2022).
\newblock {\em SAGE Open\/} .

\end{thebibliography}


\newpage
\end{document}